\documentclass[12pt]{article}      

\usepackage{graphicx}
\usepackage{amssymb}
\usepackage{amsmath}
\usepackage{lineno}
\usepackage{textcomp}
\usepackage{float}
\usepackage{fancyvrb}
\usepackage{caption}
\usepackage{graphicx,amsmath,amsfonts,amsthm,bbm,setspace,url}
\usepackage{natbib,verbatim,mathrsfs}
\usepackage{subcaption}
\usepackage{floatpag}
\usepackage{numprint}
\npthousandsep{,}

\advance\voffset by -0.25cm
\begin{document}

%\sloppy
\baselineskip=20pt

\begin{center}{
  \Large \bf
  Teleconnected warm and cold extremes of North American wintertime temperatures}
%Wintertime dependence between teleconnected warm and cold temperature extremes in North America}
%Wintertime dependence between warm Alaska extremes and cold Midwest extremes}

\bigskip

{\bf 
  Mitchell L. Krock\footnote[1]{Department of Statistics,
  Rutgers University. \\
  \quad \quad Corresponding author e-mail:
  \texttt{mk1867@stat.rutgers.edu}},
  %Author e-mail:  
  Adam H. Monahan\footnote[2]{School of Earth and Ocean Sciences, University of Victoria, Victoria, British Columbia, Canada},
  Michael L. Stein\footnotemark[1]
}

\bigskip

{\bf \today}

\end{center}

% \begin{abstract}
%   Copulas are useful in many areas, one of which being the study of extreme values. 
%   Most common models for statistical extremes are concerned with a univariate response, and copulas provide a simple way to incorporate dependence between variables whose marginal distributions are fit with classical extremes methodologies. 
%   Researchers have proposed measures of tail dependence to discern the joint behavior of variables in the tails of their distributions. 
%   In the case of bivariate copulas, tail dependence is usually focused on the ``major-diagonal''--that is, when either both variables are simultaneously far in their upper tail or far in their lower tail. 
%   Motivated by the polar vortex phenomenon in North America, where very low temperatures in the Midwest coincide with very high temperatures in Alaska, we focus on ``opposite-tail" dependence, i.e., when one variable is far in its lower tail while the other is far in its upper tail. 
%   We provide evidence of this phenomenon in reanalysis data from ERA5. An important question in extremes is thus how to construct a bivariate copula which has flexible tail behavior in all four corners of the unit square. 
%   We propose a simple solution where a mixture of rotations of common Archimedean copulas can capture various combinations of four-corner tail dependence.
%   \bigskip

  \begin{abstract}
  Current models for spatial extremes are concerned with the joint upper (or lower) tail of the distribution at two or more locations.
  Such models cannot account for teleconnection patterns of two-meter surface air temperature ($T_{2m}$) in North America, where very low temperatures in the contiguous Unites States (CONUS) may coincide with very high temperatures in Alaska in the wintertime. This dependence between warm and cold extremes motivates the need for a model with opposite-tail dependence in spatial extremes.
  This work develops a statistical modeling framework which has flexible behavior in all four pairings of high and low extremes at pairs of locations.
  In particular, we use a mixture of rotations of common Archimedean copulas to capture various combinations of four-corner tail dependence.
  We study teleconnected $T_{2m}$ extremes using ERA5 reanalysis of daily average two-meter temperature during the boreal winter. 
  The estimated mixture model quantifies the strength of opposite-tail dependence between warm temperatures in Alaska and cold temperatures in the midlatitudes of North America, as well as the reverse pattern.  These dependence patterns are shown to correspond to blocked and zonal patterns of mid-tropospheric flow.
This analysis extends the classical notion of correlation-based teleconnections to considering dependence in higher quantiles.

  \bigskip

\end{abstract}

%%%%%%%%%%%%%%%%%%%%%%%%%%%%%%%%%%%%%%%%%%%%%%%%%%%%%%%%%%%%%%%%%%%%%%%%%%%%%%%%
%%%%%%%%%%%%%%%%%%%%%%%%%%%%%%%%%%%%%%%%%%%%%%%%%%%%%%%%%%%%%%%%%%%%%%%%%%%%%%%%
%\linenumbers

\section{Introduction}
\label{sec:introduction}

% Spatial extremes is an exciting and challenging area of research which continues to garner interest. 
% Despite a large background of influential textbooks in both spatial statistics \citep{stein1999, cressie1993, wikle} and extremes \citep{coles, Gumbel+1958,Leadbetter.etal1983,GVK010852999,Haan.Ferreira2010}, there are no seminal textbooks for spatial extremes, and the field remains relatively uncharted in comparison to both research areas it combines.
% A main reason for this relative lack in progress is because the premier model in geostatistics---the Gaussian Process---is not suited for modeling extremes, as the underlying Gaussian distribution possesses a light tail and thus underestimates the possibilty of extreme events.
% Nonetheless, interest in spatial extremes has grown in recent years as environmental threats become more common and more severe in the face of climate change.

Extreme cold winter temperatures pose a large risk to human health and infrastructure.
Recently, the North American cold wave in February 2021 caused hundreds of deaths and hundreds of billions of dollars in damages after the Texas power grid failed \citep{busby2021}.
While investigation of cold extremes in a warming climate may seem counter-intuitive,  accelerated warming in the Arctic could be causing more severe winter cold weather in the midlatitudes of North America \citep{vihma2014, overland2015, perlwitz2015}.
%Since anamolously cool temperatures in the midlatitudes of the Northern Hemisphere are associated with warm temperatures in the Arctic, 
In general, cold extreme events will not be of continental scale.  Atmospheric teleconnection structures  associated with opposite-tail dependence at remote locations can result in anomalously cold air in one region being associated with anomalously warm air elsewhere. However, such teleconnection patterns have traditionally been characterized by correlations rather than extremal dependence. Teleconnections have received some attention in geostatistics \citep{hewitt2018} but not in the spatial extremes literature.
%understanding the relationship between teleconnected temperature extremes is therefore an important step to help mitigate risk associated with disasters like the Texas 2021 cold spell.
%Teleconnections are largely uninvestigated in the spatial extremes literature.
%Although teleconnections are a fundamental part of climate science and meterology, traditional teleconnection analyses focus on correlations rather than tail dependence.

A key challenge of modeling multivariate extremes  is describing the behavior of the joint distribution when multiple variables are extreme; i.e.,\ simultaneously in the tails of their respective marginal distributions.
In the context of spatial data, this poses the issue of specifying models which can transition between close-range asymptotic dependence, where nearby locations experience rare events at the same time, and long-range asymptotic independence, where distant locations experience rare events independently of one another.
Much of the current literature in spatial extremes deals with constructing models that can transition from asymptotic dependence to asymptotic independence \citep{wadsworth2012,davison2013,huser2017,huser2019}.
Such research naturally focuses on asymptotic dependence in the sense of two random variables (here, a stochastic process at two  locations) which are both in their upper marginal tails \emph{or} both in their lower marginal tails. %, with the random variables being members of a stochastic process at nearby locations.
We stress the importance of the ``or'' in the previous sentence, as methods for extremes typically only focus on a single tail, even in the univariate case. Moreover, spatial extremes models consider monotonic transitions from nearby same-tail dependence to distant independence, and do not allow for remote alternations between same- and opposite-tail dependence as might result from wavelike teleconnection structures.
%Moreover, spatial extremes models are restricted to tail dependence where all variables are either simultaneously large or simultaneously small. 
The aim of this paper is to expand the analysis of spatial extremes to accommodate teleconnection structures, as illustrated by an environmental phenomenon which exhibits 
%an interesting alternative
opposite-tail dependence where one variable is in its upper marginal tail and the other in its lower marginal tail.
Unlike the typical notion of tail dependence with spatial data, this opposite-tail dependence occurs 
%at very far 
remotely, at continental-scale distances between observation locations.
We propose a simple statistical model with the goal of modeling asymmetric tail dependencies in the four tails of a bivariate random vector.
We do not attempt to model this opposite-tail dependence within a spatial field (i.e.,\ explicitly modeling the dependence between $d > 2$ random variables) in this paper, but it remains an interesting avenue for future research.  Note that traditional, correlation-based teleconnection structures are also based on bivariate analyses.

We study opposite-tail dependence between daily boreal wintertime 2-meter temperature ($T_{2m}$) in the far Northwest of North America (Alaska) and the rest of the CONUS and Canada.
Because we are considering continental-scale extremes, we expect the the extreme events to be associated with particular anomalous circulation patterns. Previous studies have shown that topographic blocking events, associated with persistent high-pressure ridges over the subarctic Northeast Pacific, also coincide with anomalous warmth in Alaska and cold in much of the rest of the continent \citep[e.g.][]{carrera2004,jeong2021}.
%stronger type of dependence arises from 
%undulations in the position of the polar front bring warm air masses poleward and cold air masses equatorward.
%This disruption of the usual zonal flow of the polar front is often associated with a stationary high pressure region over the Northeast Subarctic Pacific which blocks the usual atmospheric circulation and persists for long periods of time (up to weeks).
%A tropospheric blocking over Alaska produces extremely cold temperatures in the midlatitudes concurrently with extremely warm temperatures in NW North America, prompting the notion of opposite-tail dependence \citep{carrera2004}.
These incidents are often called ``polar vortex'' events in media reports \citep{manney2022}. %, but we will avoid this phrase to prevent confusion and instead refer to this type of extreme event as a ``blocking'' extreme.
The other class of teleconnected $T_{2m}$ extremes centered on Alaska---namely, very cold temperatures in NW North America at the same time as very warm temperatures in the midlatitudes---has received less attention. %are less well studied.
%involves a reduction in the strength of the climatological ridge over western North America with unusually cold air in Alaska and warm air in the central part of the continent.
%We use the phrase ``zonal'' extreme in this case since the typical westerly zonal flow of atmospheric circulation is undisturbed.
% The significance of zonal extremes is likely less than blocking extremes for two primary reasons: first, very cold temperatures in the midlatitudes are more dangerous than very warm temperatures during the wintertime, and second, the inversion to warm winter temperatures in the Arctic is much more striking from a climatological perspective than the same in the midlatitudes.
% Nonetheless, comparing blocking and zonal extremes is a major part of our upcoming analysis of teleconnected $T_{2m}$ extremes.

Temperature extremes in North America have been extensively studied; see \citet{grotjahn2016} for a review.
\citet{carrera2004} examined Alaskan blocking events and found that, conditioned on the presence of blocking, there was an elevated probability of very low quantile $T_{2m}$ events in much of CONUS and an elevated probability of very high quantile events in Alaska.
 \citet{trigo2004} also studied reanalysis products to analyze blocking events over Europe.
\citet{pfahl2012} showed that warm extremes in the high latitudes are associated with co-located blocking while cold extremes are not.
\citet{xie2017} analyzed extreme cold waves over North America, particularly how they relate to circulation patterns and climate modes.
\citet{jeong2021} investigated how well climate ensembles can reproduce blocking events. 
\citet{cohen2021} and \citet{jeong2022} focused on the interaction between climate change, wintertime blocking, and cold extremes.
The effect of blocking on temperature has been studied with extreme value statistics, both over Europe \citep{buehler2011,sillman2011} and North America \citep{whan2016}.
However, these approaches employed standard univariate extremes methodology (a block-maxima approach where a generalized extreme value distribution was fitted to reanalysis and climate ensemble data) and did not examine  tail dependence coefficients as in this paper.

To study teleconnected extremes, we apply our proposed copula model to 
daily DJF (December, January, February) $T_{2m}$ across North America.
%, taken from the ERA5 reanalysis.
%reveals many intriguing spatial patterns in the estimated tail dependencies.
These estimates of tail dependence in $T_{2m}$ are further investigated by examining composite maps of geopotential height during days when teleconnected extremes occur.
We show that %the strongest measurements of teleconnected extremes are associated with a high-pressure region in northwest North America which blocks the usual westerly atmospheric circulation and leads to 
simultaneous high temperatures in Alaska and cold temperatures in the midlatitudes of North America are associated with strong blocking highs in Northwest North America.
On the other hand, under zonal flow conditions central Canada tends to favor the reverse type of opposite-tail dependence where Alaska is cold while central Canada is warm.
Studying teleconnections under the lens of tail dependencies may provide a deeper understanding of teleconnections beyond the usual correlation framework.

The paper is outlined as follows. In Section \ref{sec:taildependence}, we define tail dependence and  copulas. Sections \ref{sec:spatialextremes} and \ref{sec:data} respectively introduce some basic ideas in spatial extremes and the dataset considered in this study.
%we describe tropospheric blocking and the dataset used to study it.
In Section \ref{sec:model}, we propose a simple statistical model with flexible tail dependence in all four corners of a bivariate copula, and Section \ref{sec:results} presents results from fitting the model to the  surface air temperature from reanalysis.
We conclude in Section \ref{sec:discussion}.

\section{Tail Dependence and Copulas}
\label{sec:taildependence}

Consider two continuous random variables $X$ and $Y$ with cumulative distribution functions $F_X$ and $F_Y$. 
From Sklar's Theorem \citep{sklar1959fonctions}, the joint distribution function $F_{X,Y}(x,y)$ is equivalent to $C(F_X(x),F_Y(y))$, where the copula $C:[0,1]^2 \to [0,1]$ is a unique distribution function with uniform marginals.
Copulas are ubiquitous in the study of extremes because they offer a way to model multivariate dependence while preserving the marginal distributions $F_X$ and $F_Y$.
There are many well-known univariate extremes models which can be used to fit these marginal distributions; see Section \ref{sec:marginalmodel} for more discussion.

Upper tail dependence can be quantified by defining the tail dependence coefficient 
\begin{equation}
\label{eq:uppertaildependence}
    \lambda_{UU} = \lim \limits_{u \uparrow 1} P(F_X(X) > u \mid F_Y(Y) > u),
\end{equation} which is a measure of the probability that $X$ exceeds a value far in its upper marginal tail given that $Y$ is similarly large \citep{sibuya1960}.
Nonzero values for $\lambda_{UU}$ are categorized as ``asymptotic dependence'' while a value of zero indicates ``asymptotic independence'' and that such a conditional exceedance is improbable.
%Specifically, asymptotic independence corresponds to $\lambda_U = 0$ and asymptotic dependence corresponds to $\lambda_U \in (0,1]$.
Note that we can rewrite $\lambda_{UU}$ in terms of the copula as $\lambda_{UU} =  \lim \limits_{u \uparrow 1} (1 - 2u + C(u,u))/(1-u)$, where the numerator is the joint probability that $F_X(X)$ and $F_Y(Y)$ are larger than $u$ and the denominator is the marginal probability of a uniform random variable being larger than $u$.
%An analogous concept is $\lambda_{LL} = \lim \limits_{u \downarrow 0} P(F_X(x) \le u \mid F_Y(y) \le u) = \lim \limits_{u \downarrow  0} \frac{C(u,u)}{u}$ to assess lower tail dependence.
An analogous concept is 
\begin{equation}
\label{eq:lowertaildependence}
    \lambda_{LL} = \lim \limits_{u \uparrow 1} P(F_X(X) \le 1-u \mid F_Y(Y) \le 1-u) = \lim \limits_{u \uparrow  1} \frac{C(1-u,1-u)}{1-u}
\end{equation} to assess lower tail dependence.

\begin{figure}
  \centering
  \includegraphics[scale=.8]{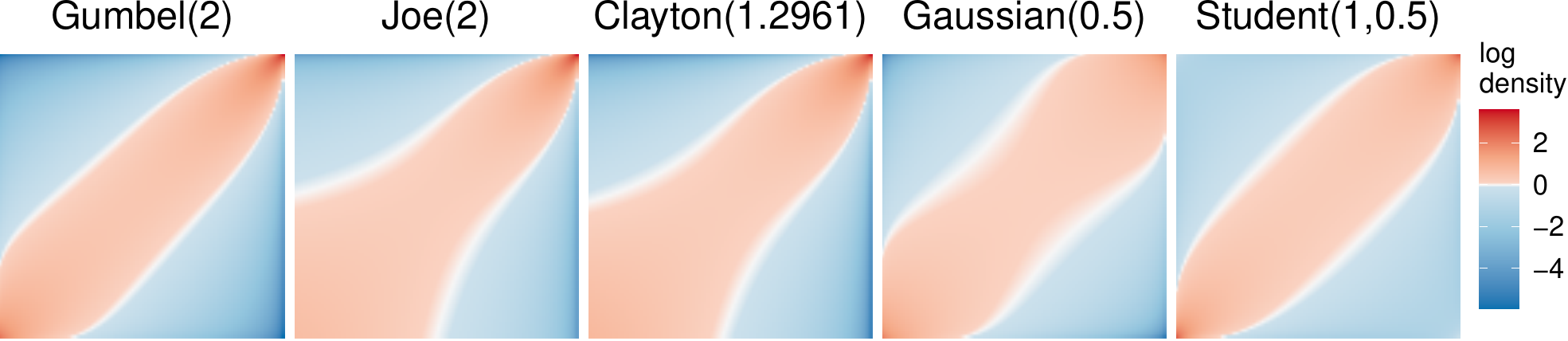}
  \caption{Log densities of five standard copulas. Axes (suppressed) are the interval $[0,1]$. White on the color scale represents a density of one; if the variables were independent, the  density would equal one everywhere on the unit square. The Clayton copula is rotated by $180^\circ$, and its parameter is chosen to give the same strength of tail dependence as the Gumbel and Joe copulas.}
  \label{fig:basiccopulas}
\end{figure}

Most standard references for copulas \citep{Nels06,Joe14} tabulate $\lambda_{UU}$ and $\lambda_{LL}$ for common copulas (e.g.,\ Archimedean and elliptical families).
These values are critical in the study of spatial extremes, as a desirable property in a model for spatial extremes is a transition from asymptotic dependence between nearby locations to asymptotic independence between distant locations.
However, the concept of opposite tail dependence has received little attention in extremes literature, and, to our knowledge, none at all in the spatial extremes literature.
With a focus on financial modeling, \citet{zhang2008} formally introduce the idea of ``total tail dependence'' for a bivariate copula.
In addition to $\lambda_{UU}$ and $\lambda_{LL}$, define the tail dependence statistics $\lambda_{UL}$ and $\lambda_{LU}$ as follows:
\begin{equation} \label{eq:oppositetaildependencies}
% \lambda_{UU} &=  \lim \limits_{u \uparrow 1} P(F_X(X) > u \mid F_Y(Y) > u) = \lim \limits_{u \uparrow 1} \frac{1-2u + C(u,u) }{1 - u}  \\
% \lambda_{LL} &=  \lim \limits_{u \downarrow 0} P(F_X(X) \le u \mid F_Y(Y) \le u) = \lim \limits_{u \downarrow  0} \frac{C(u,u)}{u} \\
\begin{split}
\lambda_{LU} &=  \lim \limits_{u \uparrow 1} P(F_X(X) \le 1-u \mid F_Y(Y) > u) = \lim \limits_{u \uparrow 1} \frac{1-u - C(1-u,u) }{1 - u},  \\
\lambda_{UL} &=  \lim \limits_{u \uparrow 1} P(F_X(X) > u \mid F_Y(Y) \le 1-u) = \lim \limits_{u \uparrow 1} \frac{1-u - C(u,1-u) }{1 - u}.
\end{split}
\end{equation}
\citet{zhang2008} also defines the tail dependence matrix 
\begin{equation}
    \label{eq:taildependencematrix}
    \begin{pmatrix}
    \lambda_{LU} & \lambda_{UU} \\
    \lambda_{LL} & \lambda_{UL}
    \end{pmatrix}
\end{equation}
as a way to summarize the tail dependencies of the bivariate random vector $(X,Y)$.  The elements of this matrix have been arranged in analogy with standard graphical representations of copulas (e.g.,\ upper tail dependence in the upper right corner, cf.\ Figure \ref{fig:basiccopulas}).
Note that the event of interest and the conditioning event can be
exchanged in any of these four tail dependencies\footnote{For example, $\lambda_{UU}$ in Eqn.\ \eqref{eq:uppertaildependence} can also be defined as $\lim \limits_{u \uparrow 1} P(F_Y(Y) > u \mid F_X(X) > u)$.} since each individual event occurs with probability $1-u$.
However, interchanging $F_X(X)$ and $F_Y(Y)$ in \eqref{eq:oppositetaildependencies} effectively switches between $\lambda_{UL}$ and $\lambda_{LU}$, so it is important to distinguish between $X$ and $Y$ for opposite-tail dependencies.
\citet{Flores2010NEGATIVEA}  studied the four corners of the bivariate copula and derived  tail dependencies for some standard copulas, which we report in Table \ref{tab:taildeptable}.
Example log densities of five standard copulas are shown in Figure \ref{fig:basiccopulas}.
Observe that none of the copulas in Table \ref{tab:taildeptable} admit different tail dependencies for the lower-right and upper-left corners of the unit square.
Indeed, no standard copula can achieve our goal of modeling different tail dependencies in the four corners of the unit square.

\begin{table}[t] 
  \tabcolsep=0.1cm
  \centering
  \begin{tabular}{c|cccc|c}
      & $\lambda_{UU}$ & $\lambda_{LL}$ & $\lambda_{UL}$ & $\lambda_{LU}$ & Parameters \\ \hline 
  Gumbel($\theta$) & $2-2^{1/\theta}$ & 0 & 0 & 0 & $\theta \in [1,\infty)$ \\
  Joe($\theta$) & $2-2^{1/\theta}$ & 0 & 0 & 0 & $\theta \in [1,\infty)$ \\
  Clayton($\theta$) & 0 & $2^{-1/\theta}$ & 0 & 0 & $\theta \in [-1,\infty) \setminus \{0\}$ \\
  Gaussian$(\rho)$ & $\mathbf{1}(\rho=-1)$ & $\mathbf{1}(\rho=-1)$ & $\mathbf{1}(\rho=1)$ & $\mathbf{1}(\rho=1)$ & $\rho \in (-1,1)$ \\
  Student($\nu,\rho)$ & $f(\nu,\rho)$ & $f(\nu,\rho)$ & $f(\nu,-\rho)$ & $f(\nu,-\rho)$ & $\nu > 0$, $\rho \in (-1,1)$
  \end{tabular}
  \caption{Tail dependencies for common copulas. Here $\mathbf{1}$ is an indicator function and $f(\nu,\rho) = 2 T_{\nu+1} \left( - \sqrt{ \frac{ (\nu+1)(1-\rho)}{1+\rho} } \right)$ where $T_{\nu}$ is the student-$t$ cumulative distribution function with $\nu$ degrees of freedom.}
  \label{tab:taildeptable}
  \end{table}

\section{Spatial Extremes}
\label{sec:spatialextremes}
Here we describe some basic aspects of spatial extremes as they relate to our goal of modeling opposite-tail dependence; see \citet{huser2022} for a recent comprehensive overview.
A current focus of spatial extremes is constructing models with common-tail asymptotic dependence between nearby locations which can smoothly transition to asymptotic independence as the distance between locations grows larger.
Opposite-tail dependencies $\lambda_{LU}$ and $\lambda_{UL}$ are unstudied in the spatial extremes literature, as values at two nearby locations are generally positively correlated and highly unlikely to be in opposite-tails at the same time.
However, classical correlation-based teleconnection patterns such as the North Atlantic Oscillation describe spatial processes in which remote points are negatively correlated. 
%Teleconnection structures corresponding to dependence of extremes have received little attention in the literature.
%A main purpose of this paper is to provide evidence of opposite-tail dependence occurring in the environment, suggesting that a desirable property for spatial extremes models is the ability to describe negatively-dependent teleconnected extremes. 
We both provide evidence of the occurrence of opposite-tail dependence in the environment and describe an approach to modeling this behavior statistically.  Note that wavelike teleconnection structures could result in the recurrence of same-tail dependence at points more remote than those displaying opposite-tail dependence.  We do not explicitly investigate this possibility in this study, although it is admitted by our approach to modeling bivariate extremes.
%, which will hopefully spur researchers in spatial extremes to think more broadly about tail dependence instead of concentrating solely on $\lambda_{UU}$ and $\lambda_{LL}$.
%In particular, we study the ``polar vortex'' phenomenon, which produces very warm temperatures in Alaska concurrently with very cold temperatures in the Midwest of the United States during winter months.

Standard methodology for spatial extremes can be viewed as an extension of common techniques for univariate extremes to multiple dimensions.
See \citet{coles} for an introduction to extreme value statistics. 
The bulk of the distribution is typically ignored in models for extremes, as including non-extreme observations in inference can introduce bias into the fit in the tails.
With spatial extremes, extant models only consider the case when all variables are far in the upper tail of their marginal distribution.
The lower tail of a joint distribution can be modeled (separately from the upper tail) in the same manner as the upper tail after multiplying the data by $-1$.

Multivariate max-stable distributions are the generalization of the univariate generalized extreme value distribution to multiple dimensions.
For spatial data, this framework models the pointwise block-maxima of independent realizations of a stochastic process at different locations.
Inference for multivariate max-stable distributions is extremely computationally challenging, although recent progress has been made in this direction \citep{lenzi2021, husermaxstable, hector2022}.
More fundamentally for the present setting, max-stable processes are not suited for the problem of simultaneous extremes.

Peaks-over-threshold modeling can also be generalized to a spatial process.
For univariate extremes, a generalized Pareto distribution is used to model a random variable conditional on it being larger than a pre-specified threshold.
In the extension to a spatial process, the conditioning event is a cost or risk functional\footnote{Examples of such cost/risk functionals of a standardized spatial process $Y(\mathbf{s})$ include $\{ \sup \limits_{\mathbf{s} \in \mathcal{S}} Y(\mathbf{s}) \}$ or $\int_{\mathcal{S}} Y(\mathbf{s}) d \mathbf{s}$ where $\mathcal{S}$ is a spatial domain and $\mathbf{s}$ is a location in $\mathcal{S}$.} of the (standardized) spatial process. 
Inference for generalized Pareto processes is also very difficult, although not as innately intractable as likelihood calculations for max-stable processes.

Despite the prevalence of max-stable and generalized Pareto processes, there is no immediate way to extend them to model opposite-tail dependence while retaining the theoretical justification for using these methods.
Specifically, to model opposite-tail dependence, the cost functional for generalized Pareto processes would need to balance between the stochastic process being large and small at different locations.
Research in spatial extremes has begun to turn away from generalized Pareto processes because of their inability to transition from asymptotic dependence to asymptotic independence, but they seem to have a more fundamental restriction when considering opposite-tail dependence. 
Recently developed
conditional spatial extremes models \citep{wadsworth2022} also lack the ability to model opposite-tail dependence in teleconnected extremes.
Random scale constructions \citep{huser2017, huang2019, engelke2019} can exhibit opposite-tail dependence, as seen in the simple case of a student-$t$ copula in Table \ref{tab:taildeptable}.
However, they encounter challenges when faced with asymmetric distributions, nonstationary patterns of tail dependence, and changing asymptotic dependence classes over space, all of which are present in our data analysis.
As a result of these limitations, in this study we consider the more modest (but tractable) question of modeling bivariate extremal dependence in a spatial field.
%We now turn our attention to  an example of a meteorological phenomenon resulting in  opposite-tail dependence in surface air temperature.

%\section{Atmospheric Blocking and Data}
\section{Data}
\label{sec:data}

\label{sec:data}
We use reanalysis products from ERA5 \citep{ERA5} (see \url{https://cds.climate.copernicus.eu/cdsapp#!/dataset/reanalysis-era5-single-levels?tab=overview}) to study teleconnected $T_{2m}$ extremes.
ERA5 provides hourly two-meter temperature across the globe on a 0.25 degree longitude-latitude grid, available from  1979 to mid-2022.
Recently, ERA5 extended records back to 1959, but we avoid using measurements from the pre-satellite era for this analysis.
We examine the daily average two-meter temperature in the boreal winter months of December, January, and February. %(including the end of February 2022).
In total, there are 3940 values of daily average temperature at each location.
Analysis is restricted to Canada and the U.S.\ to constrain the study.  Daily-mean 500 hPa geopotential heights from ERA5 are also used to produce circulation composites associated with joint temperature extreme events.
%We focus on North America to constrain the study, although research shows that the polar vortex has started to shift towards the Eurasian CONUS in recent decades \citep{zhang2016}.

As extreme values are likely oversmooothed in reanalysis, working with in-situ measurements would be preferred, but observational data pose issues in the form of record quality, missing data, and the spatial consistency of the measurements. 
Area-averaging should reduce problems in extremes with reanalysis data and represent true areal averages more accurately than reanalysis gridpoints represent station data.
We accordingly divide the domain into $4^{\circ}$ latitude $\times$ $6^{\circ}$ longitude gridboxes within which $T_{2m}$ data are averaged over space.
This spatial averaging alleviates the computational burden of fitting models to thousands of sites and also removes some of the small-scale variability associated with individual locations.
Estimated opposite-tail dependencies between pairs of gridboxes from our coarse regional grid were also observed to be stronger than those between pairs  from the local reanalysis resolution grid (not shown).
The gridbox size was selected to make the gridboxes in the range of latitudes we consider approximately square in shape.
Gridboxes are slightly offset from integer valued longitude/latitude lines so that reanalysis gridpoints are not grouped in multiple coarse gridboxes.
Finally, we make an arbitrary choice to keep only gridboxes where greater than 85\% of observations are over land. 
Within the 76 selected gridboxes, only the observations over land are averaged to create a temperature record of 3940 days for each gridbox.
Example plots of the study region with selected gridboxes are shown in Section \ref{sec:results}.

\section{Statistical Model}
\label{sec:model}

Statistical modeling of the temperature data follows a two step approach.
First, a parametric  model with flexible behavior in both tails is used to model the marginal distribution of temperature at each location.
This marginal fit takes into account both seasonality and climate change; see Section \ref{sec:marginalmodel} and Appendix \ref{app:bats} for more details.
With a model for the marginal distribution at each location, the observations are transformed to uniformity by applying the estimated cumulative distribution function. %\footnote{This transformation to uniformity (c.f.\ ``probability integral transform'') refers to the fact that $F_X(X)$ follows a standard uniform distribution for any random variable $X$.}.
Then, dependence between pairs of uniform random variables is modeled with a mixture copula (Section \ref{sec:copulamodel}).
Parameter estimation for both steps, along with the transformation to uniformity, is described in Section \ref{sec:mle}.

\subsection{Marginal Model}
\label{sec:marginalmodel}

Since our goal is a single model with flexible behavior in all four corners of a bivariate copula, we need a marginal model that can simultaneously fit the upper and lower tails of a univariate distribution.
To this end, we use a recently proposed a parametric distribution with flexible behavior in both tails \citep{steinbats}.
We refer to this seven-parameter model as BATs (``Bulk-And-Tails'') and emphasize that this is a departure from standard univariate extremes methodologies (i.e.,\ block-maxima and peaks-over-threshold approaches) which only consider a single tail of the distribution at a time and also ignore the bulk of the distribution.
The BATs cumulative distribution function equals $T_\nu(H_\theta(x))$ where $T_{\nu}$ is the student-$t$ cumulative distribution function with $\nu$ degrees of freedom and $H_\theta$ is a monotone increasing function depending on the other six parameters.
Allowing parameters of the BATs model to depend on covariates can represent nonstationary behavior such as seasonality and climate change \citep{krock2022}.
Appendix \ref{app:bats} provides more detail about the nonstationary marginal fits.
After the marginal models are specified, data are marginally transformed to uniformity, and we turn to copulas to model the dependence between pairs of uniform random variables.
Results presented in subsequent sections are similar under different transformations to marginal uniformity, such as the empirical distribution function\footnote{Given independent and identically distributed realizations $x_1,\dots,x_m$ of a random variable $X$, the empirical distribution function is given by $\hat{F}_X(x) = \frac{1}{m} \sum_{i=1}^m \mathbf{1}[x_i \le x]$.} instead of the estimated Bulk-And-Tails distribution function.

\subsection{Copula Model}
\label{sec:copulamodel}
To model the four-corner tail dependence of a bivariate copula, we propose a mixture of four copulas, each with distinct tail dependence in a separate corner. 
This can be achieved by mixing four rotations of a copula which has tail dependence in only a single corner of the unit square. 
By rotating a bivariate copula, we mean reorienting the copula by $90^\circ, 180^\circ,$ or $270^\circ$ (e.g.,\ Figure \ref{fig:basiccopulas} shows a Clayton copula rotated by $180^\circ$). %, which is done in practice by applying the mapping $u \mapsto 1-u$ to different combinations of the two inputs to the copula.
Options for copulas with tail dependence only in a single corner of the unit square are common members of the Archimedean family such as the Gumbel, Joe, and Clayton copulas.
The student-$t$ copula exhibits tail dependence in all four corners of the unit square, but it is symmetric in opposite-tails, and therefore cannot model different tail dependencies in the four corners of the unit square even with a mixture of rotations.

Let $C_\theta(u_1,u_2)$ be a bivariate copula with tail dependence in a single corner, and let $c_\theta(u_1,u_2) = \frac{\partial^2}{\partial u_1 \partial u_2} C_\theta(u_1,u_2)$ be its density.
We model dependence between uniform random variables $U_1$ and $U_2$ with the copula
\begin{equation} \label{eq:mixture}
  P(U_1 \le u_1, U_2 \le u_2) =  \sum_{k=1}^4 w_k C_{\theta_k}(u_1,u_2)
\end{equation}
where $w_1,\dots,w_4$ are nonnegative weights with $\sum_{k=1}^4 w_k = 1$ and $C_{\theta_k}$ are four copulas which are rotated to capture four different tail dependencies.
Let $\lambda_{LL,k}, \lambda_{UL,k}, \lambda_{LU,k}$, and $\lambda_{UU,k}$ denote the four-corner tail dependencies of the copula $C_{\theta_k}$ for $k=1,\dots,4$.
Then, since $C_{\theta_k}$ has tail dependence matrix $M_k = \begin{pmatrix} \lambda_{LU,k} & \lambda_{UU,k} \\ \lambda_{LL,k} & \lambda_{UL,k} \end{pmatrix}$, the copula mixture \eqref{eq:mixture} has tail dependence matrix $\sum_{k=1}^4 w_k M_k$ \citep{zhang2008}.
By forcing each $M_k$ to have only one nonzero entry in a different corner of a $2 \times 2$ matrix, we are able to obtain different tail dependencies in the four corners of the copula.
A limitation of our proposed mixture model is that the sum of the four-corner tail dependencies is less than or equal to one, which can be problematic at nearby locations when $\lambda_{UU}$ and $\lambda_{LL}$ may both be both large.
See Appendix \ref{app:tailconstraint} for a mathematical explanation of this restriction.

Some previous works have considered rotating copulas for different tail dependencies, mainly in the context of financial time series. % \citep{zhang2008,NING2009202,ROSSI201394,chang2021}.
\citet{zhang2008} used rotations of a bivariate Gaussian copula in their data analysis, which is ineffective in our purpose because the Gaussian copula only has tail dependence in the case of perfect correlation.
\citet{NING2009202} and \citet{ROSSI201394} rotated copulas to model different upper and lower tail dependencies $\lambda_{UU}$ and $\lambda_{LL}$.
\citet{gong2022asymmetric} construct a copula model with flexible, asymmetric behavior in the joint upper and joint lower tails.
\citet{chang2021} proposed a model very similar to our mixture of rotated Gumbel copulas but with a slightly different weight function. \citet{chang2021} also uses a GARCH model for the marginal time series distributions.
Most importantly, \citet{chang2021} models dependence on returns in financial instruments and does not consider spatial dependence.
To the best of our knowledge, our work is the first to consider opposite-tail dependence in a spatial setting.

\subsection{Estimation of model parameters}
\label{sec:mle}
Both the marginal and copula distributions are fit via maximum likelihood while assuming independence across days in the likelihood function, and the two estimation steps are performed separately.
The independence assumption should not bias point estimates of parameters, but temporal dependence must be taken into account during uncertainty quantification.
The practice of fitting marginal distributions, transforming each random variable to uniformity, and then fitting a copula to model dependence on the unit hypercube is common in copula literature and known as ``inference for margins'' (IFM).
IFM is much easier in terms of the optimization than joint estimation of marginals and copulas, and, under some regularity conditions, IFM estimates are consistent and asymptotically normal \citep{joe2005}.
In  our case, a full optimization for the marginals and copula at the same time was found to be ineffective unless IFM solutions were used as the initial parameter guess, and, even with a good initial guess, the fitting was still time consuming and encountered issues with convergence.
Separating the parameter inference for marginals and copula will also help us more clearly discern the effects of different copula models and different choices made about aggregating the temperature data.

For a given gridbox with $m=3940$ daily average $T_{2m}$ values $x_1,\dots,x_m$, the marginal BATs loglikelihood reads
\begin{equation} \label{eq:marginalloglikelihood}
  \mathcal{L}_{m} = \sum_{i=1}^m \left[\log \left( t_\nu(H_\theta(x_i)) \right) + \log (H'_\theta(x_i))\right]
\end{equation}
where $t_\nu$ is the student-$t$ density with $\nu$ degrees of freedom and $H'_\theta$ is the derivative of $H_\theta$.
Parameters of the BATs distribution vary with time to capture seasonality and climate change.  Appendix \ref{app:bats} provides more details about the parameterization and code for fitting the model.
The estimated BATs cumulative distribution function is applied to each daily $T_{2m}$ value to create uniformly distributed data. % $u^{(1)},\dots,u^{(m)}$.
This transformation to uniformity (cf.\ ``probability integral transform'') follows from the fact that $F_X(X)$ follows a standard uniform distribution for any random variable $X$.

For independent and identically distributed random vectors $(u_{1,1},u_{2,1}),\dots,(u_{1,m},u_{2,m})$ on the unit square, the loglikelihood of the mixture model reads
\begin{equation} \label{eq:mixtureloglikelihood}
  \mathcal{L}_{c}=\sum_{i=1}^m \log \left( \sum_{k=1}^4 w_k c_{\theta_k}(u_{1,i},u_{2,i}) \right)
  %\sum_{i=1}^m \log( w_1 c_{\theta_1}(u_1^{(i)},u_2^{(i)}) + w_2 c_{\theta_2}(1-u_1^{(i)},u_2^{(i)}) + w_3 c_{\theta_3}(1-u_1^{(i)},1-u_2^{(i)}) + w_4 c_{\theta_4}(u_1^{(i)},1-u_2^{(i)})).
\end{equation}
Both models are fit by maximum likelihood in Julia using the IPOPT solver \citep{ipopt} with automatic differentiation from the package ForwardDiff.jl \citep{forwarddiff}.
Code for fitting the marginal loglikelihood \eqref{eq:marginalloglikelihood} and the copula loglikelihood \eqref{eq:mixtureloglikelihood} is available at \texttt{github.com/mlkrock/BulkAndTails.jl} and \texttt{github.com/mlkrock/TeleExtrTemp}, respectively.
In total there are seven independent parameters for the copula model: four Archimedean copula parameters and three weights $w_1,w_2,w_3 \in [0,1]$ which are constrained so that $0 < w_1 + w_2 + w_3 < 1$.
A variety of initial guesses were provided to the optimization software if the original attempt using $w_k = 0.25$ and $\theta_k = 2$ for $k=1,\dots,4$ did not converge.
For the rest of the paper, we show results using rotations of the Gumbel copula, but any of the Archimedean copulas from Table \ref{tab:taildeptable} are reasonable choices and produced similar values in terms of estimated tail dependencies.

\section{Results}
\label{sec:results}

Here we analyze results from fitting our copula model to the 76 gridboxes of daily winter two-meter temperature.
We consider both the correlation used to define conventional teleconnection patterns and estimated tail dependencies between gridboxes.
Unlike the correlation between two random variables, which is symmetric in its two arguments, the tail dependencies between two random variables are not necessarily the same in all four corners of the unit square.
Of particular interest here are the spatial behavior of these tail dependencies, especially the opposite-tail dependencies $\lambda_{LU}$ and $\lambda_{UL}$ defined in Eqn.\ \eqref{eq:oppositetaildependencies}.
Common-tail dependencies defined in Eqns.\ \eqref{eq:uppertaildependence} and \eqref{eq:lowertaildependence} are also important, but they are a common focus of spatial extremes and generally behave in a more predictable way: $\lambda_{UU}$ and $\lambda_{LL}$ are relatively large when looking at two nearby gridboxes, and the dependence weakens as distance between the gridboxes increases.  We do not find evidence in the data considered of remote increases in $\lambda_{UU}$ or $\lambda_{LL}$ associated with distant positive teleconnection pattern centers, although this possibility cannot be excluded on physical grounds. 

\begin{figure}
  \centering
  \includegraphics[scale=.4]{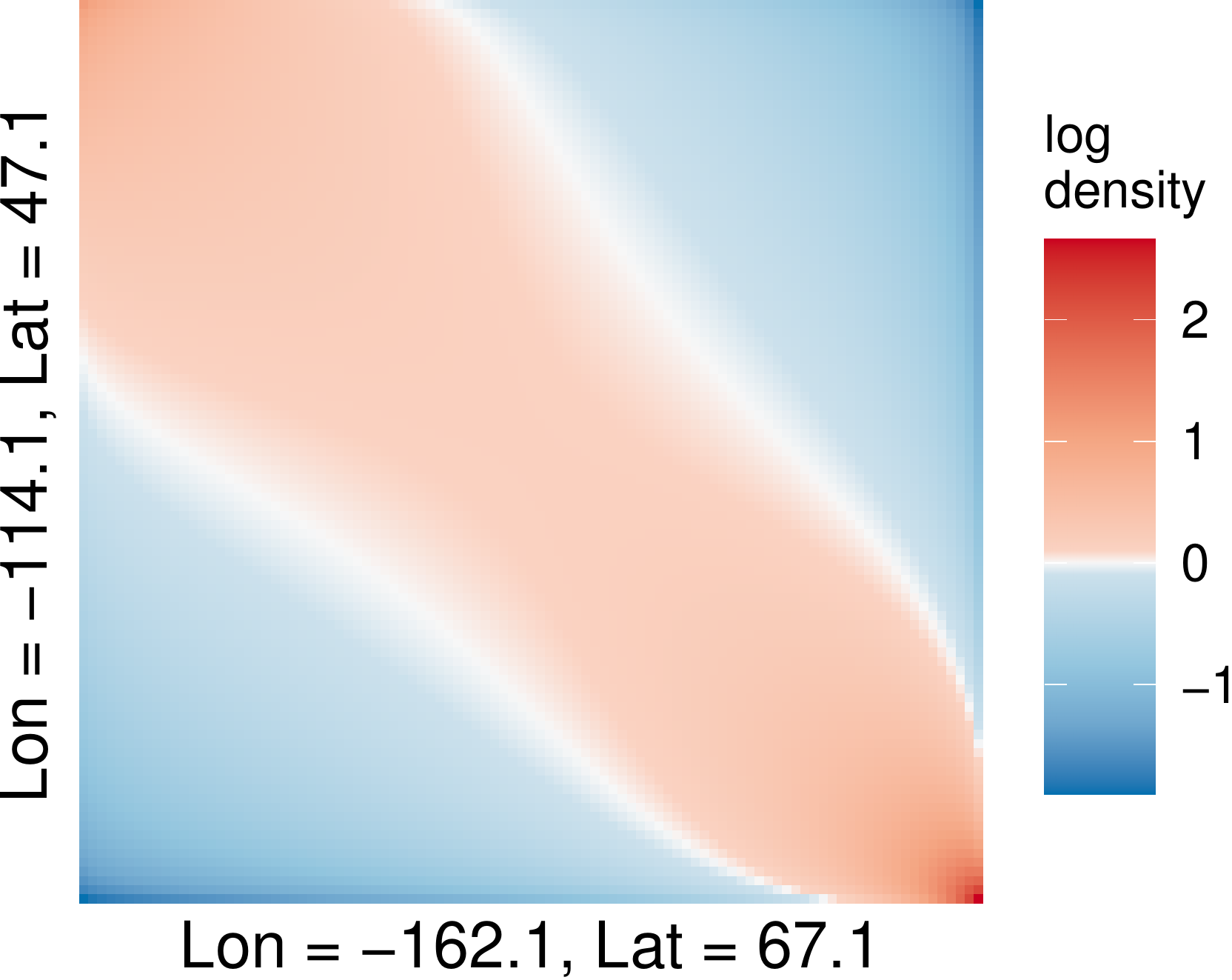}
  \includegraphics[scale=.4]{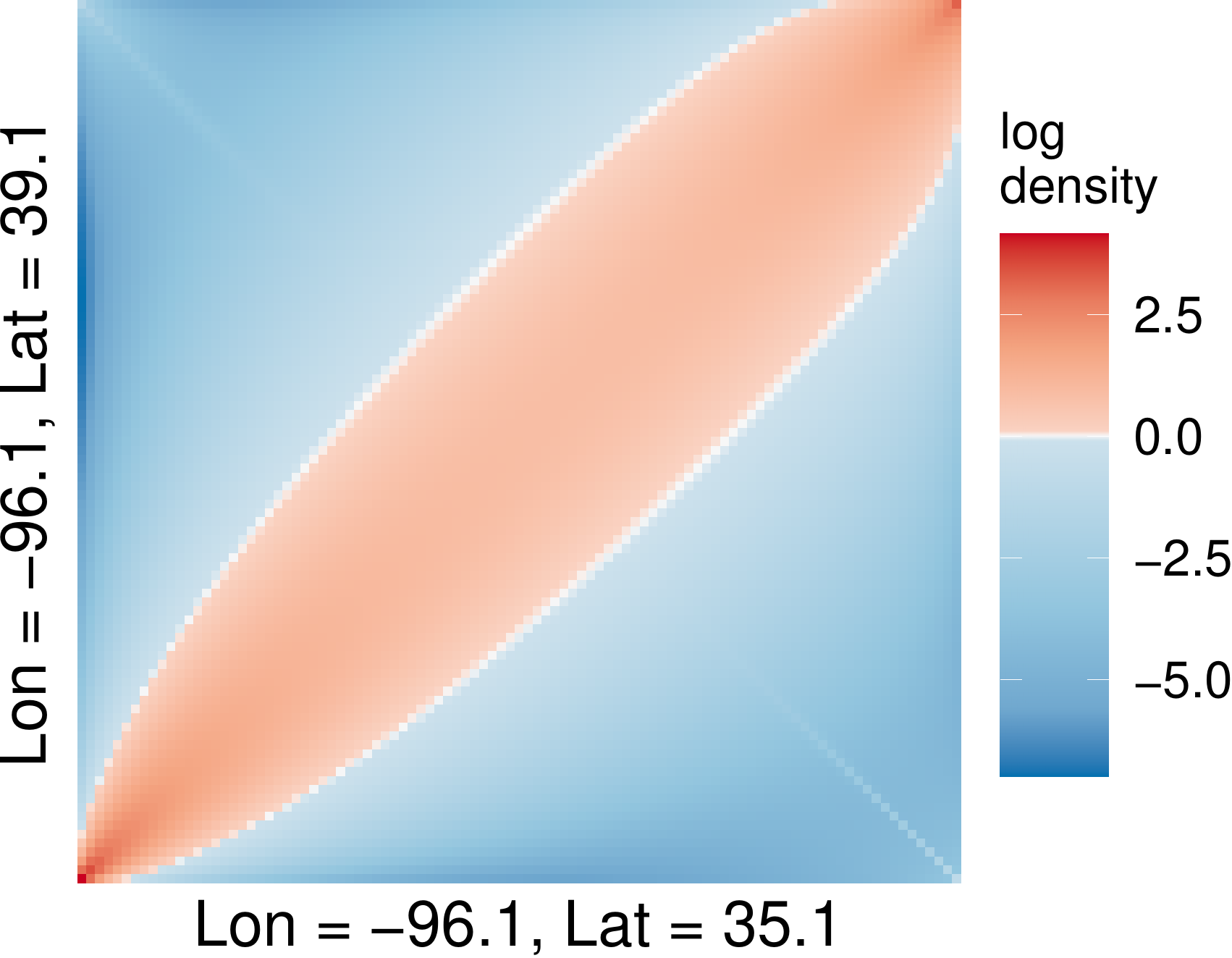}
  \caption{Estimated log copula densities at two pairs of sites. Axes (suppressed) are the interval $[0,1]$. White on the color scale represents a density of one; if the variables were independent, the  density would equal one everywhere on the unit square.}
  \label{fig:estimatedcopuladensities}
\end{figure}

As an illustration of the versatility of the rotated copula mixture model, we plot examples of estimated log copula densities in Figure \ref{fig:estimatedcopuladensities}.
The left and right plots respectively show log copula densities for the pairs of gridboxes with the largest opposite-tail dependence and the largest lower tail dependence.
Specifically, the left plot shows the density where one gridbox is in northwest Alaska and the other is in Montana (the corresponding latitudes and longitudes are noted on the axes).
%The density indicates a fairly long range negative dependence between surface air temperature in Alaska and Montana, which could be caused by behaviors in the jet stream.
The density displays a striking asymmetry in the upper-left and lower-right corners in both the bulk of the density and in the corners.
There is a larger area with density greater than one in the upper left region than in the lower right region of the unit square, yet the density in the lower right corner is much greater than in the upper left corner.  %, as would be expected from a blocking extreme event where the Alaska gridbox is warm while the Montana gridbox is cold.
In contrast, the right plot shows results for two adjacent gridboxes in southern-central U.S.,\ where temperatures are strongly positively correlated.
Upon close inspection, there is a similar, albeit less pronounced, asymmetry along the main diagonal: the upper right region of the unit square has more area with density larger than one, yet the density is largest in the lower left corner of the unit square.
In some other cases, it can be difficult to visually discern differences in tail dependence statistics from plots of the copula density.
Some additional density plots shown on the temperature scale are available in the Supplementary Material.
%Importantly, the mixture model is able to produce densities which behave like rotations of the Archimedean copulas in Figure \ref{fig:basiccopulas} without the need to specify whether the dependence is positive or negative or where it is strongest.  

\begin{figure}
  \centering
  \includegraphics[scale=.6]{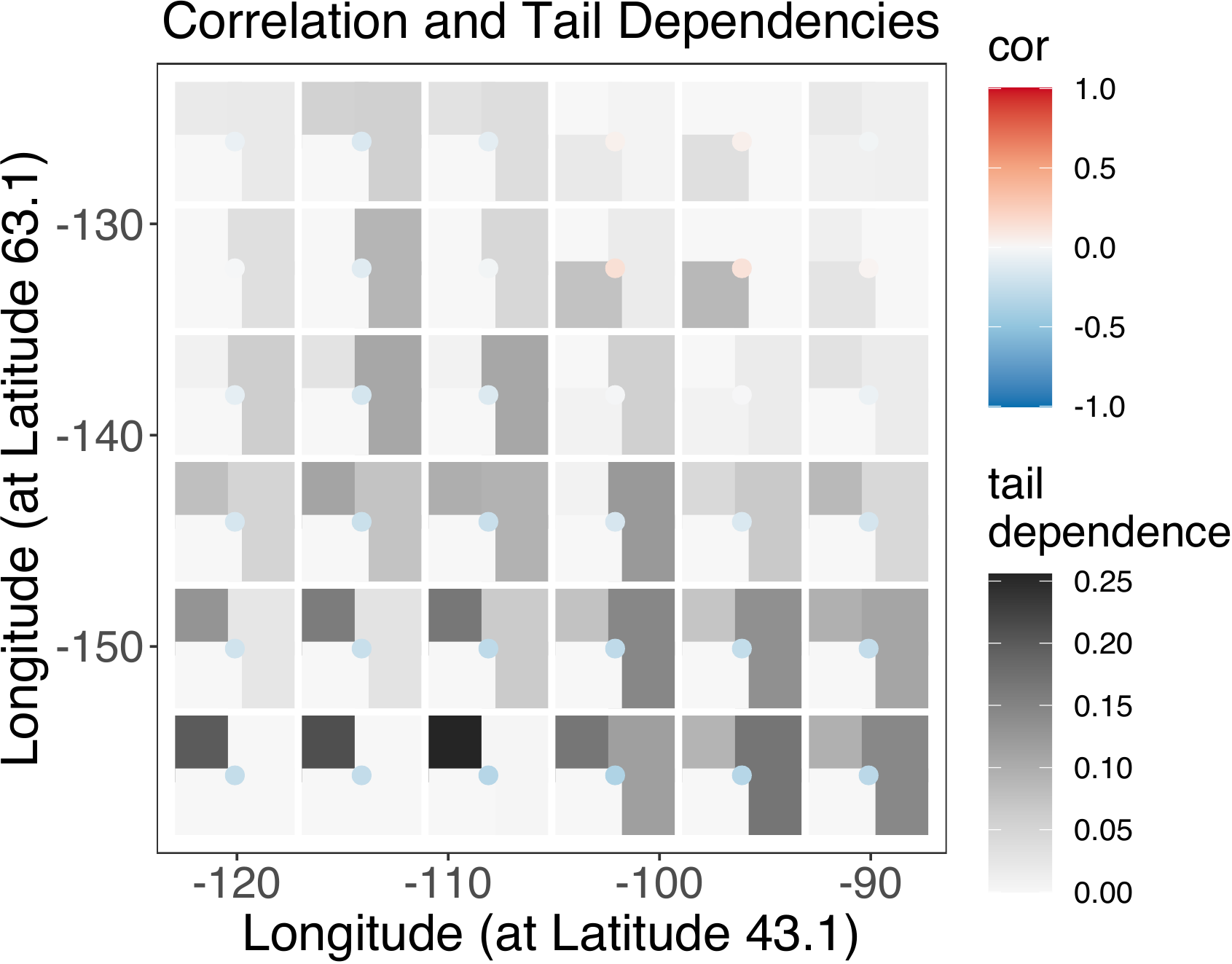}
    \includegraphics[scale=.23]{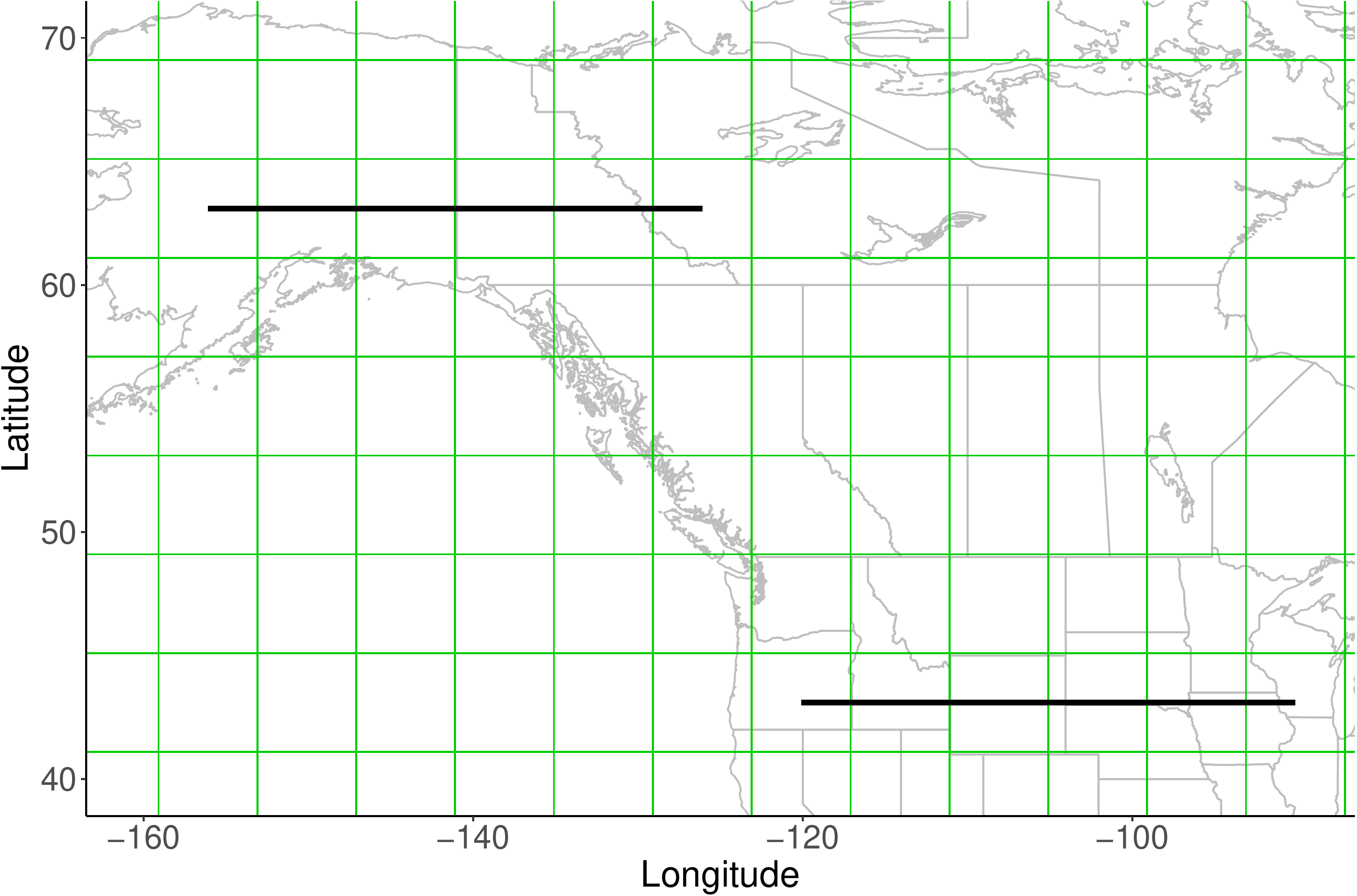}

  \caption{Tail dependencies and correlation between pairs of gridboxes along two bands of fixed latitude.
  Each colored correlation point is centered at a gridbox's centroid, and the grayscale depicts the tail dependence matrix defined in Eqn.\ \eqref{eq:taildependencematrix} with the southern latitude band (i.e.,\ $x$-axis) corresponding to the $X$ variable in Eqn.\ \eqref{eq:oppositetaildependencies}. As a reminder, $\lambda_{LU}$ refers to the upper-left quadrant of the four corners.  The green graticule in the right-hand panel denotes the spatial grid used for averaging.}
  \label{fig:twobandtaildependencies}
\end{figure}

Figure \ref{fig:twobandtaildependencies} shows the spatial evolution of tail dependencies and correlation between pairs of gridboxes at two fixed latitude bands,  one  starting in Alaska and the other in Oregon.
All correlations are relatively weak between these pairs of stations, but negative correlations are stronger than positive ones. % between the lower latitude band and the western gridboxes of the upper latitude band.
Shaded grayscale squares in the larger boxes show the values of the tail dependence matrix from Eqn.\ \eqref{eq:taildependencematrix}.
Tail dependencies are strongest between the westernmost gridboxes; i.e.,\ the strongest tail dependencies are in the $\lambda_{LU}$ direction of anomalously high temperatures in the northern latitude band and anomalously low temperatures in the southern latitude band.
The other three tail dependencies between these western gridboxes are relatively weak, and $\lambda_{UL}$ becomes largest for easternmost gridboxes along the southern latitude band. %as the southern latitude gridbox moves to the east.
Joint positive tail dependence $\lambda_{UU}$ is largest when the gridbox in either the northern or southern latitude band is relatively far east while the other is more westward. %, which is expected due to the natural eastward movement of weather systems, yet $\lambda_{UU}$ is largest when the upper gridbox is far to the west, while $\lambda_{LL}$ is largest when the upper gridbox is far to the east. 
Joint negative tail dependence $\lambda_{LL}$ is only noticeable in the small region where correlation is slightly positive, while fairly large values for $\lambda_{UU}$ appear even when correlation is negative.  These results illustrate how different the spatial structure of the four measures of tail dependence can be from each other and from the correlation coefficient.

\begin{figure}
  \centering
  \includegraphics[scale=.5]{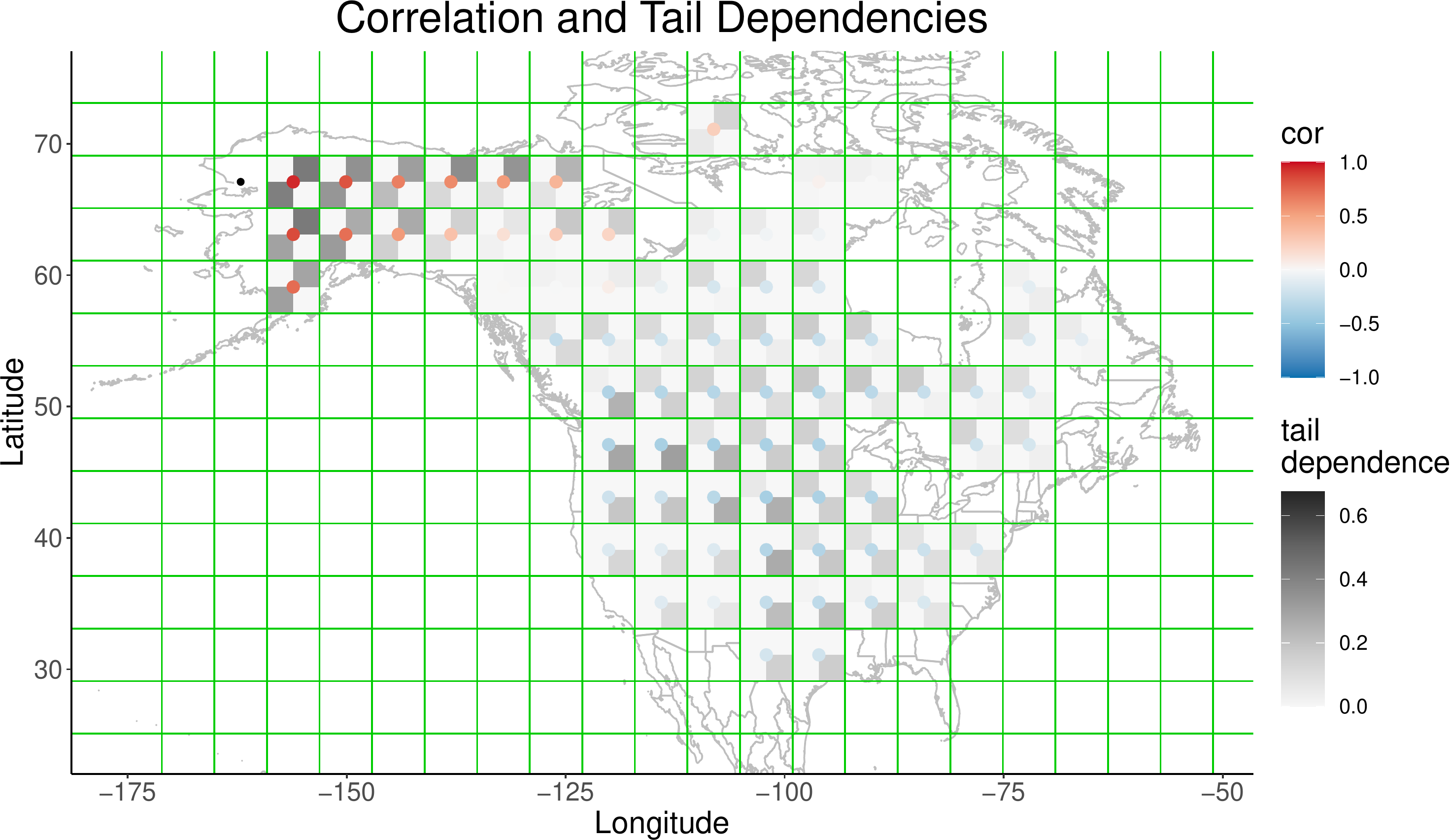}
  \caption{Correlation and estimated tail dependencies between gridbox in northwest Alaska (black dot) and all other pairs of gridboxes. For reference, the black dot  corresponds to a gridbox in northwest Alaska whose lower left coordinate is $(-165.1^\circ, 65.1^\circ)$ and upper right coordinate is $(-159.1^\circ, 69.1^\circ)$.
    Grayscale denotes the gridbox values of the extremal dependence matrix (such that the lower right corner of a gridbox depicts $\lambda_{UL}$, the tail dependence where that gridbox is especially cold and the gridbox with the black dot is especially warm.)  The coarse gridboxes are delimited by the green graticule.}
  \label{fig:maptaildependencies}
\end{figure}

Displaying tail dependencies and correlation for a single base point as a spatial map provides another perspective on the spatial structure of extremal dependence. %showcases many interesting characteristics of $T_{2m}$. % and the best fit model.
Figure \ref{fig:maptaildependencies} shows such a map of dependence between a central gridbox in Alaska and all other gridboxes.
A full animation changing the central gridbox is available in the Supplementary Material and \texttt{github.com/mlkrock/TeleExtrTemp}.
The map shows an interesting transition from positive dependence in Alaska and northwest Canada, through no dependence in northern Canada, to negative dependence in southern Canada and CONUS.
We start by examining the region of negative dependence, as it is the physically the largest.
%and its importance is the crux of this paper.
Negative correlation with the central point in Alaska is present in all gridboxes lower than roughly 55$^{\circ}$N  and is strongest in southern Canada, the northern U.S.,\ and the Midwest.
Opposite-tail dependencies are the strongest overall when the northwest CONUS is cold and Alaska is warm.

A noteworthy feature in this region of negative dependence is the spatial relationship between $\lambda_{LU}$ and $\lambda_{UL}$.
In the gridboxes in Canada, $\lambda_{LU}$ (cold-Alaska to warm-Canada tail dependence) is larger than $\lambda_{UL}$, while in the United States, $\lambda_{UL}$ (warm-Alaska to cold-CONUS tail dependence) is larger than $\lambda_{LU}$ and is particularly large in the northwest U.S. and the Midwest.
We emphasize that these opposite-tail dependencies exhibit noticeably different spatial structures despite having similar correlations with the central gridbox.
In the region where spatial dependence is weak, correlations as well as tail dependencies are near zero.
Finally, even the region of positive dependence exhibits interesting tail asymmetries, as values directly east from the central gridbox are larger in $\lambda_{UU}$ than $\lambda_{LL}.$

\begin{figure}
  \centering
  \includegraphics[scale=.5]{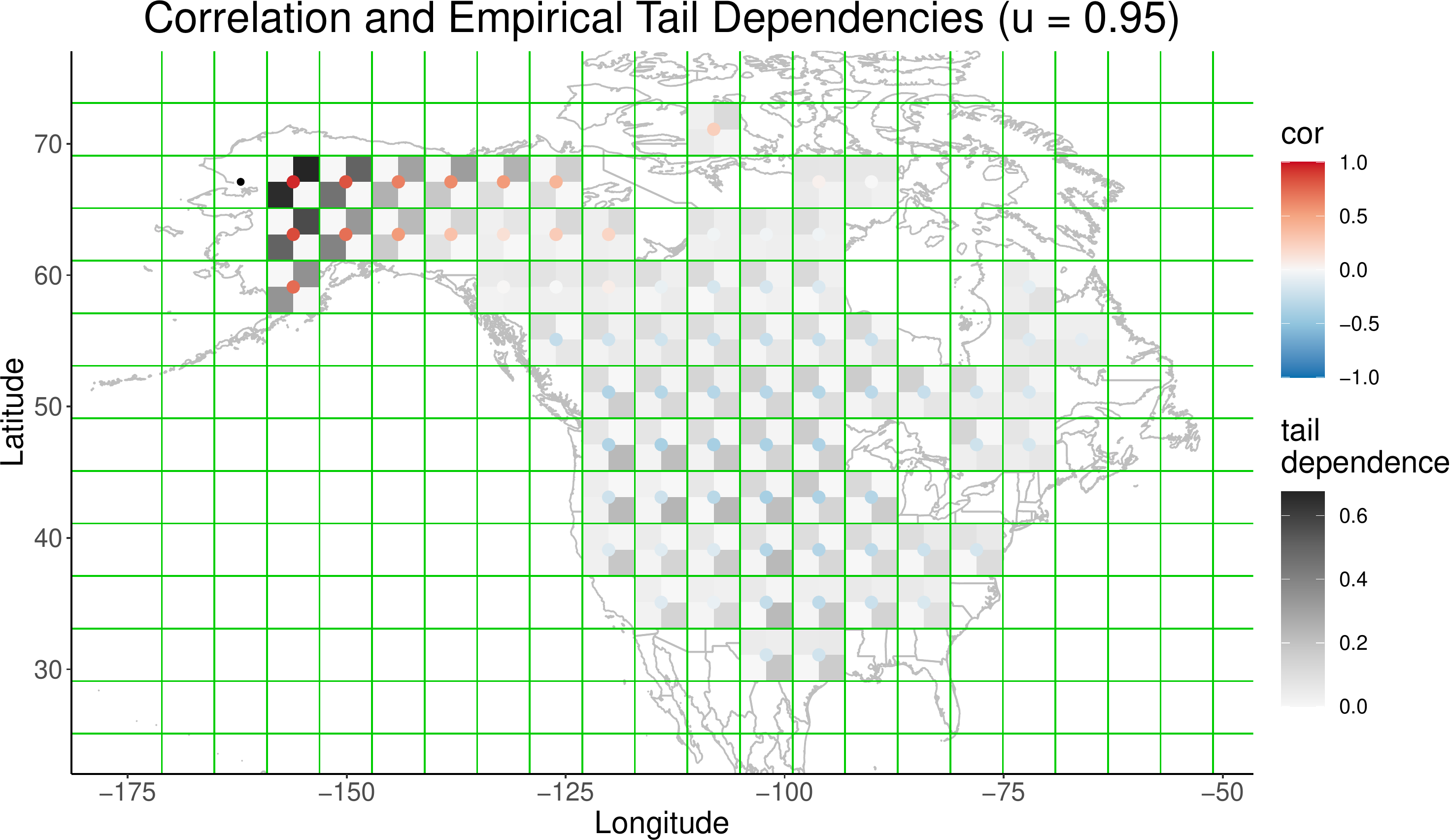}
  \caption{As in  Figure \ref{fig:maptaildependencies} with empirical estimates of the tail dependencies based on counts of simultaneous quantile exceedance. A threshold quantile of $u=0.95$ is used in this case.}
  \label{fig:mapemmpiricaltaildependencies}
\end{figure}

So far, we have %presumed the existence of 
used the parametric model fit to investigate extremal dependence. % due to abnormalities in the polar front.
To %check if such an assumption is reasonable, 
check the robustness of these results, we can examine empirical estimates of tail dependencies obtained from raw counts of simultaneous quantile exceedances.  
Section \ref{app:empiricaltaildependence} in the Appendix gives the formulas for these empirical estimates.
Figure \ref{fig:mapemmpiricaltaildependencies} displays the analogous empirical estimates of the tail dependencies shown in Figure \ref{fig:maptaildependencies} using a threshold  of $u=0.95$.
Figure \ref{fig:mapemmpiricaltaildependencies} shows clear evidence of dependence in moderately extreme outcomes that qualitatively matches the spatial patterns seen in Figure \ref{fig:maptaildependencies}.
As mentioned in Section \ref{sec:copulamodel}, the mixture model has a limitation in that the sum of the values of its tail dependence matrix cannot be greater than one.
Indeed, empirical estimates of $\lambda_{UU}$ and $\lambda_{LL}$ at gridboxes near the black dot are noticeably larger than the corresponding parametric estimates.
To facilitate comparison, tail dependence values in Figure \ref{fig:mapemmpiricaltaildependencies} share the same scale as those in Figure \ref{fig:maptaildependencies}, but empirical estimates of $\lambda_{UU}$ and $\lambda_{LL}$ can exceed this scale, and those which do so are colored black.
Empirical estimates of tail dependence are quite unstable if the thresholds are substantially increased from $u=0.95$.

\begin{figure}
  \centering
  \includegraphics[scale=.5]{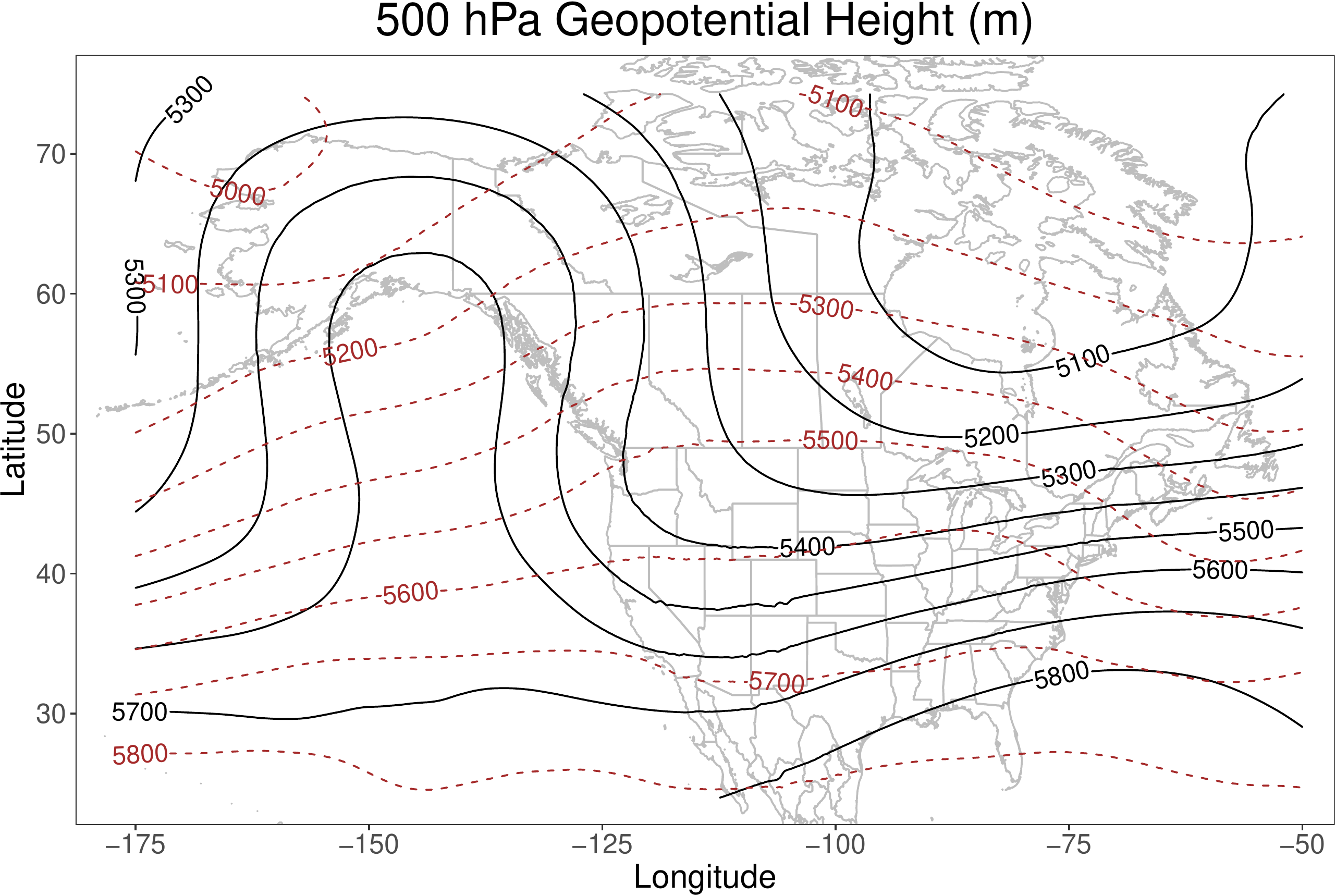}
  \caption{Contours of 500 hPA geopotential height (in meters) composited over days with simultaneous $T_{2m}$ threshold exceedances (cf.\ Figure \ref{fig:mapemmpiricaltaildependencies} and Appendix B). Black contours correspond to warm-Alaska-cold-CONUS blocking extremes, while brown contours are for cold-Alaska-warm-Canada zonal extremes. The gridpoints used to determine simultaneous extreme exceedances are noted in the text.}
  \label{fig:geopotential}
\end{figure}

Examining composite circulation patterns during opposite-tail extreme events can provide additional insight into negatively dependent teleconnections in North America.
Figure \ref{fig:geopotential} shows contour plots of the daily average geopotential height at 500 hPa, composited over days with extreme $T_{2m}$ teleconnections which are defined empirically as in Appendix \ref{app:empiricaltaildependence} and Figure \ref{fig:mapemmpiricaltaildependencies} with a value of $u = 0.95$.
Warm-Alaska-cold-CONUS extremes are calculated with respect to the black-dot gridbox in Alaska and the gridbox centered at coordinates $(-114.1^{\circ},47.1^{\circ})$, and for the cold-Alaska-warm-Canada pattern, the second gridbox is centered at $(-90.1^{\circ},51.1^{\circ})$.
These locations correspond to the largest values for $\lambda_{UL}$ and $\lambda_{LU}$ in Figure \ref{fig:maptaildependencies}.
In total, there are 42 daily maps which are composited for the warm-Alaska-cold-CONUS extremes and 33 for the opposite case.
The geopotential height distributions clearly indicate that warm-Alaska-cold-CONUS extremes are associated with a high-pressure tropospheric blocking ridge over the Notheast Subarctic Pacific and a low pressure trough over central Canada.  Warm Pacific air is advected into Alaska while cold high-latitude air is advected into the Canadian Prairies and American Midwest.
In the cold-Alaska-warm-Canada case, the geopotential height distribution is much more zonal.  
The climatological high pressure ridge over western North America is absent and the flow over western Canada has a southerly component.  
Low pressure over Northwest Alaska results in cold air advection into Alaska, while relatively warm Pacific air is advected into the continental interior in the midlatitudes.  
These geopotential height features illustrate the role of anomalous circulation patterns in producing the teleconnected $T_{2m}$ extremes.
%the geopotential height is much smaller over Alaska and larger over Canada and nearly all of CONUS.
%This different behavior in Southern Canada is unsurprising; when the jet is anomalously zonal, Alaska should be cold and the Southern Candadian Prairies should be relatively warm.

\begin{figure}
  \centering
  \includegraphics[scale=.5]{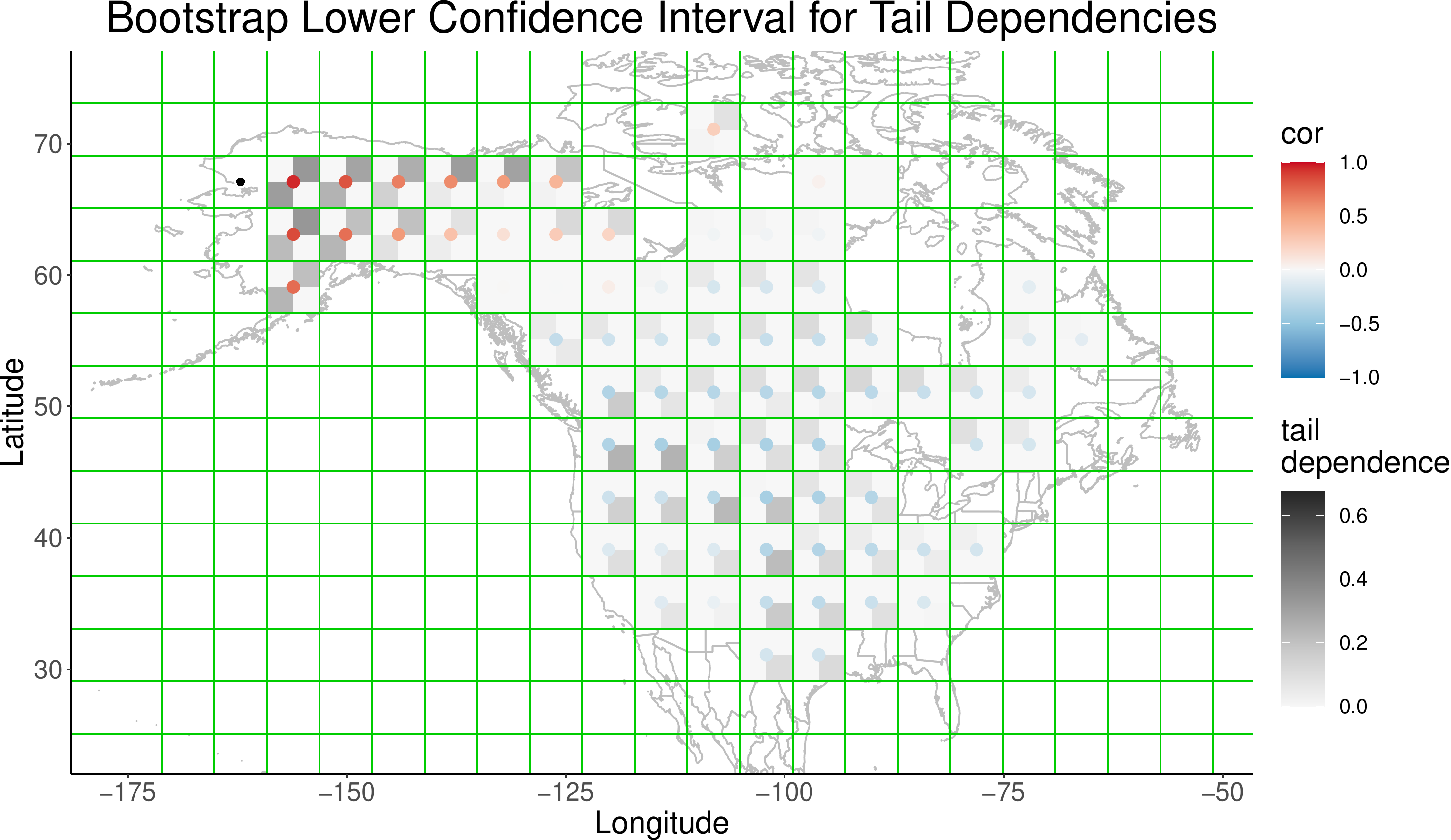}
  \caption{As in Figure \ref{fig:maptaildependencies} for the lower bound of bootstrap 95\% confidence interval for tail dependencies.} %depicted in Figure \ref{fig:maptaildependencies}.}
  \label{fig:boottaildependencies}
\end{figure}

We also explore uncertainty in tail dependencies and their spatial patterns with bootstrapping.
Although maximum likelihood estimation in Section \ref{sec:mle} assumes observations are independent, we conduct a block bootstrap procedure to account for temporal dependence in this uncertainty quantification.
Specifically, a bootstrapped dataset is created by resampling years from the historical $T_{2m}$ data while maintaining the number of years in each decade\footnote{Decade groupings were 1979-1989, 1990-1999, 2000-2009, and 2010-2022.} to preserve any patterns due to climate change.  This approach assumes that serial dependence associated with seasonality is accounted for in the model for the marginal distributions.
In total, 200 bootstrapped datasets are created, and marginal and copula parameters are estimated for each bootstrapped dataset as discussed in Section \ref{sec:mle}.
A 95\% percentile bootstrap confidence interval for the parameter $\hat p$ is given by $(\hat{p}_{2.5\%}, \hat{p}_{97.5\%})$, where $\hat{p}_{2.5\%}$ and $\hat{p}_{97.5\%}$ are the 2.5 and 97.5 percentiles of the bootstrap estimates for $\hat{p}$.
We use the percentile bootstrap instead of alternative methods such as the basic bootstrap to ensure that the confidence interval for a tail dependence coefficient lies in the interval $[0,1]$ \citep{PACIOREK201869}.
Figure \ref{fig:boottaildependencies} shows the lower bound from a 95\% bootstrap confidence interval for the tail dependence estimates in Figure \ref{fig:maptaildependencies}.
The basic spatial patterns of the estimated four tail dependencies are preserved in the bootstrap confidence intervals.
We again see strong same-tail dependence near the gridbox in Alaska, a transition zone with close to no dependence, and finally two distinct regions of zonal and blocking teleconnections in the midlatitudes.
Again, values are especially large for tail dependence associated with a warm-Alaska-cold-Plains blocking teleconnection.
The lower uncertainty bounds of opposite-tail dependencies in the midlatitudes are large enough that the modeled opposite-tail dependence seems to be a robust result.

\section{Discussion and Conclusions}
\label{sec:discussion}
Parametric modeling of opposite-tail dependence in spatial extremes has been uninvestigated until now. 
Conventional meteorological analyses of remote negative dependence have considered correlation-based teleconnection patterns. 
We have generalized the notion of teleconnections to remote extremal dependence through proposing a mixture of rotated Archimedean copulas to capture asymmetric tail dependence in the four corners of a bivariate copula.
The mixture model was fit to daily average DJF two-meter surface air temperatures in North America from the ERA5 reanalysis.
Resulting asymmetries in the estimated tail dependencies display many noteworthy spatial patterns; particularly notable is the strength of teleconnected extremes for which Northwest North America is unusually warm while the Plains of the CONUS are very cold.
Maps of geopotential height composites in Figure \ref{fig:geopotential} indicate that teleconnected extreme events correspond to different types of circulation patterns.
Specifically, the warm-Alaska-cold-CONUS extremes are associated with high-pressure blocking regions over Alaska, while cold-Alaska-warm-Canada extremes correspond to zonal flow.  The occurrence of high/low quantile $T_{2m}$ events  conditional on the existence of a Northwest Pacific blocking ridge has been demonstrated in previous studies \citep[e.g.,\ upper decile $T_{2m}$ in Alaska, lower decile $T_{2m}$ in the Plains in][]{carrera2004}. We provide a parametric approach to direct investigation of teleconnected extremes without the need for an external conditioning variable.  This approach is a first step in the generalization of spatial extreme analysis to accommodate remote opposite-tail dependence.

The analysis in Section \ref{sec:results} not only revealed the necessity of modeling opposite-tail dependence in spatial data, but also the need for opposite-tail dependence to transition between asymptotic independence and asymptotic dependence, which is already a major focus of current research in spatial extremes.  However, current approaches consider these questions only in the context of upper (lower) tail dependence.
We note that there exists a concept of tail order to gauge extremal dependence in the case of asymptotic independence \citep{ledford1996} for which opposite-tail spatial dependence is also worth considering.
Teleconnected opposite-tail dependence implies the need for an alternative to the standard approach to spatial extremes, as the opposite-tail dependence coefficient should be zero at nearby locations and take maximum values at remote locations.
In this work we only considered negative teleconnections, but positive teleconnections in the joint upper (lower) tails are also possible \citep{huang2019}.
The transition from asymptotic dependence to asymptotic independence in the joint upper (lower) tail is usually presumed to be monotonic with respect to distance, but this does not need to be the case.
For example, for teleconnection patterns taking the form of planetary-scale wave trains, $\lambda_{UU}$ and $\lambda_{LL}$ could increase again at distance after initially falling.

For our bivariate analysis, the mixture of rotated Archimedean copulas is easy to fit and provides a useful description of dependence in the four tails of the unit square.
It may be desirable to construct a model for total tail dependence that can avoid potential biases from the bulk of the distribution, which could be achieved through censoring non-extreme observations in the maximum likelihood estimation.
Although the proposed copula model possesses flexible tail dependencies, it is presumably limited for analyzing extremes in some cases, as only two parameters (a copula parameter and its weight) are used to describe each joint tail. 
As discussed in Section \ref{sec:copulamodel}, the sum of the four tail dependencies of our mixture model cannot exceed one. 
This constraint was evident when comparing empirical and parametric estimates of $\lambda_{UU}$ and $\lambda_{LL}$ at nearby gridboxes.
% The model may also be limited in its ability to provide a natural transition between asymptotic dependence and independence.
% In practice, the estimated mixture weights $w_1,\dots,w_4$ produced by  numerical optimization will be nonzero and therefore the mixture model will never truly exhibit asymptotic independence. %transition to asymptotic independence when locations are very far apart.
% However, some weights can be very small in magnitude, which effectively translates to very weak tail dependence. 

The most prominent limitation of the mixture model is in its restriction to a small number of variables.
Rotating mixtures of copulas to capture all tail dependencies between $d>2$ random variables quickly becomes untenable, as it requires rotations into $2^d$ corners of the unit hypercube and two parameters (a weight and a copula parameter) for each rotation.
Therefore, an open question is how to construct a model for spatial extremes with total tail dependence, as our proposal is not feasible even with a moderate number of spatial locations.
An important feature such a model must possess is that the tail dependence between two locations cannot effectively depend only on spatial correlation. 
Figure \ref{fig:maptaildependencies} illustrates several cases where gridboxes have similar correlation values but markedly different tail behaviors which also show spatial patterns.
In particular, CONUS and southern Canada possess similar values of negative correlation with Alaska, but the cold-Alaska-warm-Canada tail dependence is larger than the opposite direction of tail dependence in mid-southern Canada, while tropospheric blocking causes the warm-Alaska-cold-CONUS tail dependence to be exceptionally large, especially in northwestern CONUS.

The classical concept of correlation-based teleconnections has long been fundamental to the investigation of large-scale variability in the climate system.  This study provides a generalization of this idea  to allow quantification of remote dependence of extremes.  While we have considered application of this approach to extremes of surface air temperature, remote opposite-tail dependence of other meteorological fields (e.g.,\ precipitation, wind power density) is an interesting direction of future study.
% Creating a unified model with flexible behavior in all possible tails of a spatial process would be a challenging but worthy endeavor for the spatial extremes community.

\section*{Acknowledgements}
Mitchell Krock and Michael Stein acknowledge support from the U.S. Department of Energy, Office of Science, Office of Advanced Scientific Computing Research (ASCR), Contract DE-AC02-06CH11347. Adam Monahan acknowledges the support of the Natural Sciences and Engineering Research Council of Canada (NSERC) (funding reference RGPIN-2019-204986).

\section*{Data Availability Statement}
ERA5 data products are publicly available. For $T_{2m}$, refer to \url{https://cds.climate.copernicus.eu/cdsapp#!/dataset/reanalysis-era5-single-levels?tab=overview}, and for 500 hPa geopotential height, refer to \url{https://cds.climate.copernicus.eu/cdsapp#!/dataset/reanalysis-era5-pressure-levels?tab=overview}.

\bibliographystyle{rss} 
\bibliography{vortexbib}

\appendix
\section{Marginal Distribution}\label{app:bats}
The standard seven parameter Bulk-And-Tails (BATs) distribution \citep{steinbats} is a model for an entire univariate distribution with flexible behavior in the upper and lower tails. 
\citet{krock2022} extended BATs to a nonstationary distribution by allowing some parameters to depend upon time-varying covariates.
Here, we describe how we fit nonstationary marginal distributions to the reanalysis data.
The parameterization attempts to account for (wintertime) seasonality and climate change.

BATs has location parameters $\phi_i$ and scale parameters $\tau_i$ for the upper $(i=1)$ and lower $(i=0)$ tails.
Let $y$ represent the year and $d$ represent the number of days since the most recent December 1.
Suppressing the subscripts $i=0,1$ for $\tau_i$ and $\phi_i$, we use the following parameterizations:
\begin{equation}
\begin{split}
  \phi(d,y) &= \alpha_0 + \alpha_1 d + \alpha_2 d^2 +  \beta C(y) \nonumber\\ 
  \log(\tau(d)) &= \gamma_0 + \gamma_1 d + \gamma_2 d^2
  \end{split}
\end{equation}
$C(y)$ denotes the value of the log CO$_2$ equivalent at year $y$ and is used as a proxy for anthropogenic climate change due to greenhouse gas emissions. 
CO$_2$ equivalent data are obtained from (Version 2.2 of) the PRIMAP emission time series \citep{co2data}, which includes annual values up to the year 2018. 
We regress PRIMAP CO$_2$ on the historic Mauna Loa CO$_2$ dataset to produce values at years after 2018.
Code for fitting this model is available at \texttt{github.com/mlkrock/BulkAndTails.jl}.

\section{Tail Restriction of Mixture Model}
\label{app:tailconstraint}
The limitation on the sum of the tail dependence matrix of our mixture model is a byproduct of how it is constructed.
Recall that each rotated Archimedean copula only contributes nonzero tail dependence to a single, distinct corner of the unit square.
Thus, the tail dependence matrix of the mixture model equals 
$\begin{pmatrix}
w_{LU} \chi_{LU} & w_{UU} \chi_{UU} \\
w_{LL} \chi_{LL} & w_{UL} \chi_{UL}
\end{pmatrix}$
where $w_{LU}, w_{UU},
w_{LL}, w_{UL}$ are nonnegative weights adding to one and $\chi_{LU}, \chi_{UU}, \chi_{LL}, \chi_{UL}$ are the tail dependence coefficients of the appropriate rotated Archimedean copula. 
Each of these eight values are in the interval $[0,1]$. Therefore,
$
w_{LU} \chi_{LU} + w_{UU} \chi_{UU} +
w_{LL} \chi_{LL} + w_{UL} \chi_{UL} 
\le w_{LU}+ w_{UU} +
w_{LL} + w_{UL} = 1.
$

\section{Empirical Tail Dependencies}
\label{app:empiricaltaildependence}
Suppose $(x_1,y_1),\dots,(x_m,y_m)$ are independent and identically distributed realizations of $(X,Y)$.
Equation (2.62) in \citet{reiss2007} defines the  empirical estimate of $\lambda_{UU}$ as
\begin{equation*}
  \hat \lambda_{UU}(u) = \frac{1}{m(1-u)} \sum_{i=1}^m \mathbf{1} \left( x_i > x_{\lfloor um \rfloor : m} \text{ and } y_i > y_{\lfloor um \rfloor : m} \right)
\end{equation*}
where $\mathbf{1}$ is an indicator function and $x_{\lfloor um \rfloor : m}$ is the $\lfloor um \rfloor$ largest value of $(x_1,\dots,x_m)$.
Empirical estimates of $\lambda_{LL}, \lambda_{LU}$, and $\lambda_{UL}$ are defined similarly.

\end{document}

%\sloppy
\baselineskip=20pt

\begin{center}{
  \Large \bf
  Teleconnected warm and cold extremes of North American wintertime temperatures}

\bigskip

{\bf 
  Supplementary material
}

\bigskip

{\bf \today}

\end{center}

Supplementary material also includes an mp4 file. This animation corresponds to Main Figure 4 in the paper and shows the map as the central gridbox (indicated by a black dot) is changed over the 76 gridboxes.

\section{Densities}

\begin{figure}[h!]
  \centering
  \includegraphics[scale=.4]{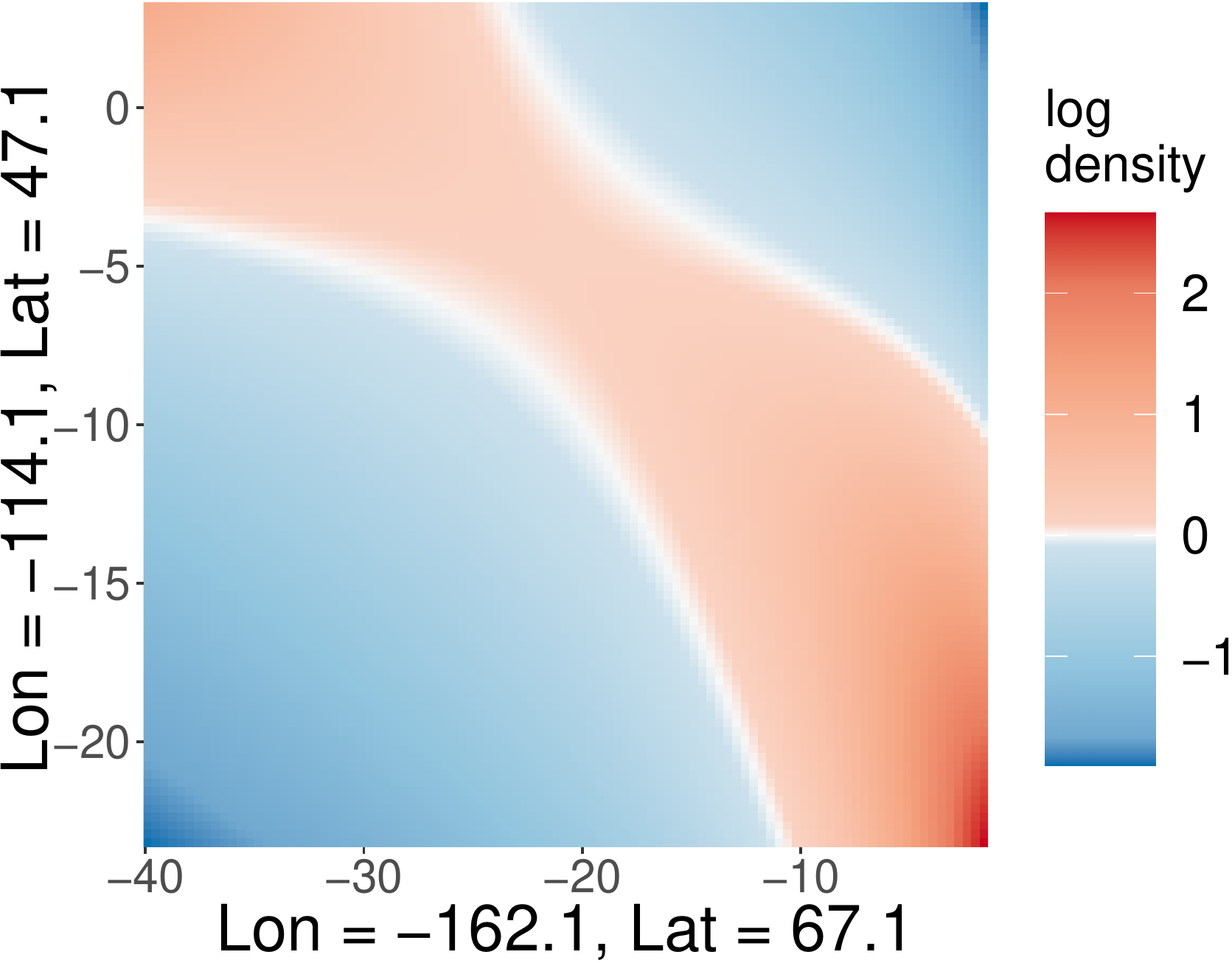}
  \includegraphics[scale=.4]{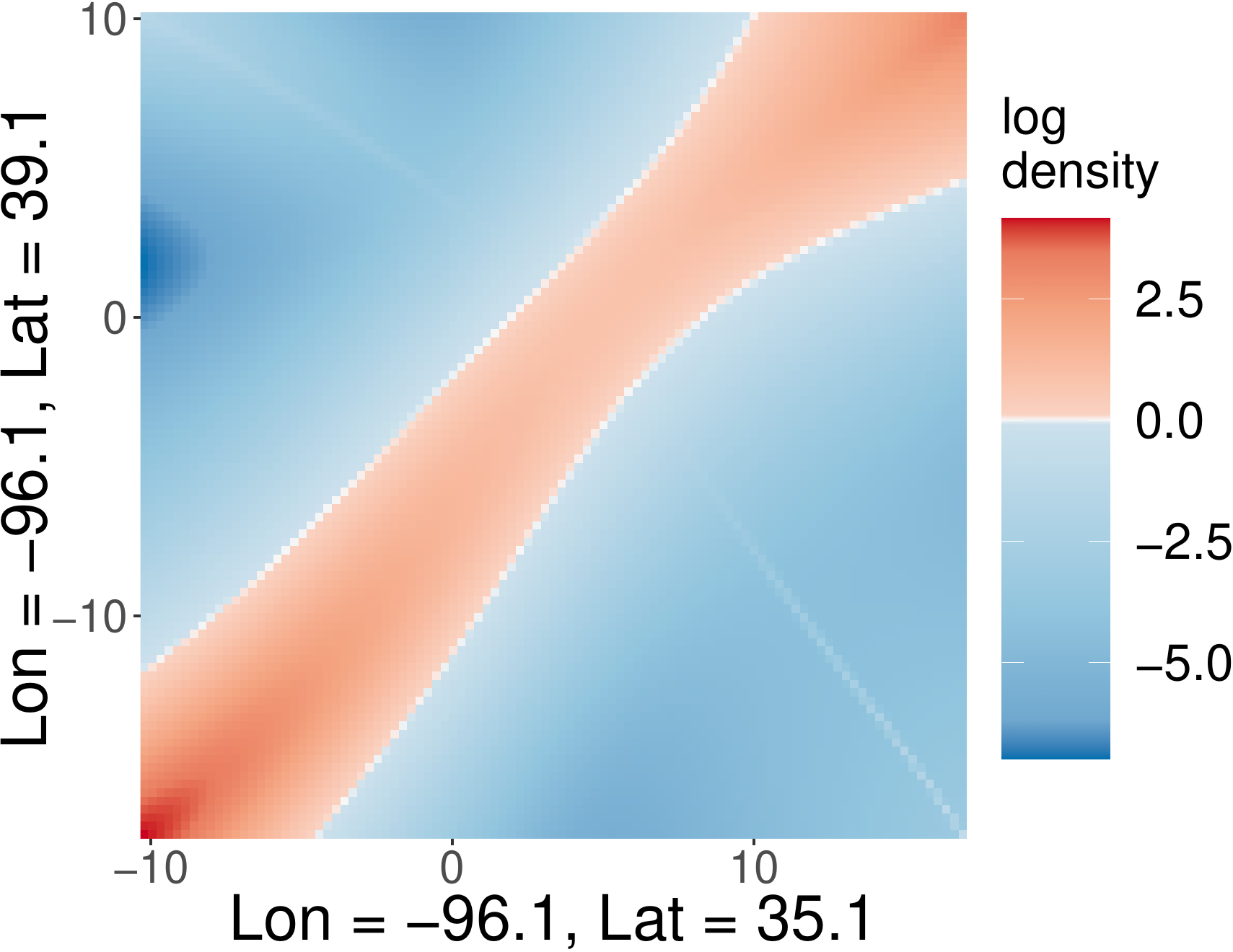}
  \caption{Estimated log copula densities at two pairs of sites. Axes are the unit interval $[0,1]$ rescaled to temperature scale (degrees Celsius). White on the color scale represents a density of one (i.e.,\ independence).}
  \label{fig:margscalelogdensity}
\end{figure}

\begin{figure}[h!]
  \centering
  \includegraphics[scale=.4]{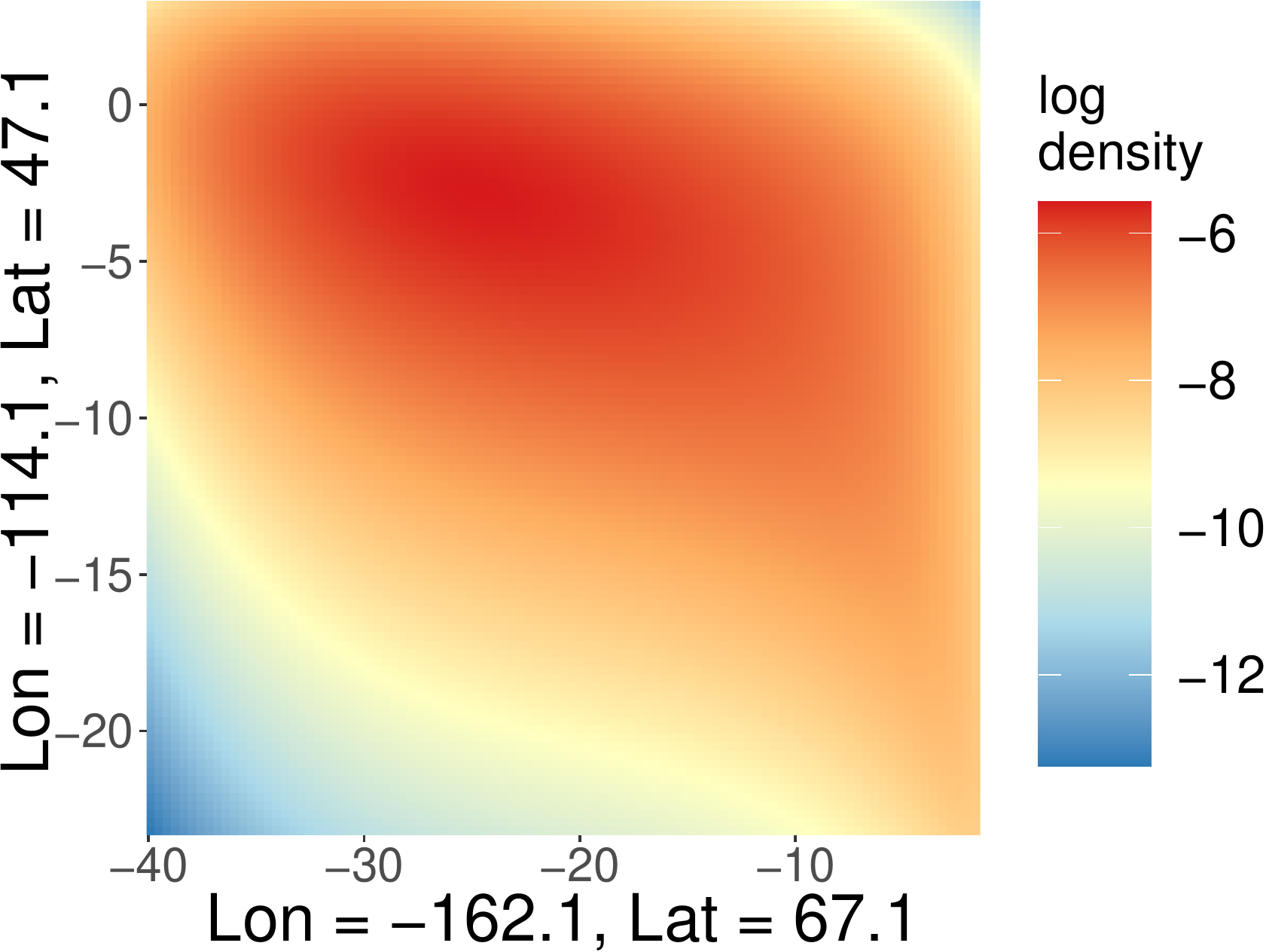}
  \includegraphics[scale=.4]{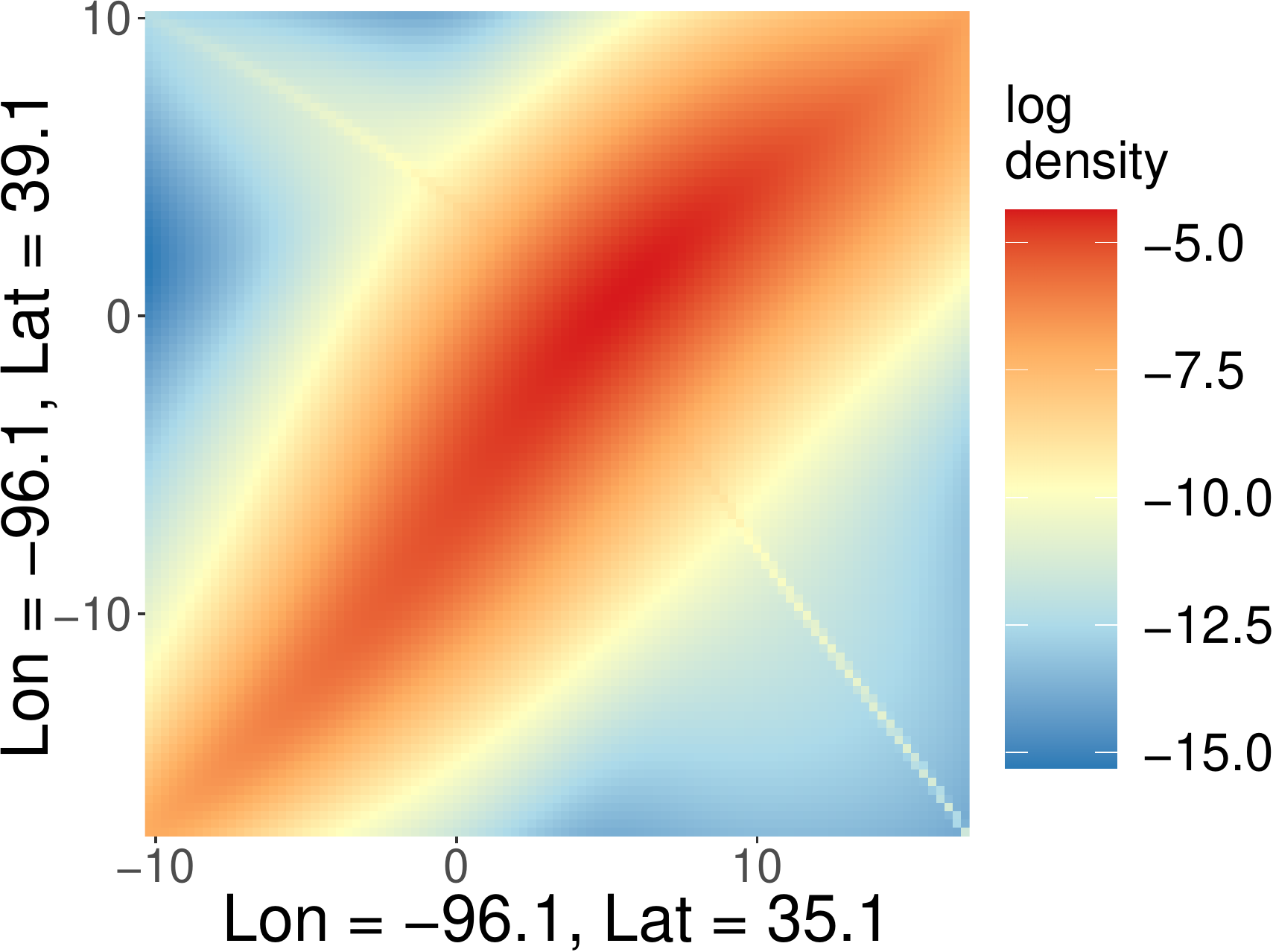}
  \caption{Estimated log joint densities at two pairs of sites. Axes are temperatures in degrees Celsius.}
  \label{fig:jointlogdensity}
\end{figure}

\begin{figure}[h!]
  \centering
  \includegraphics[scale=.4]{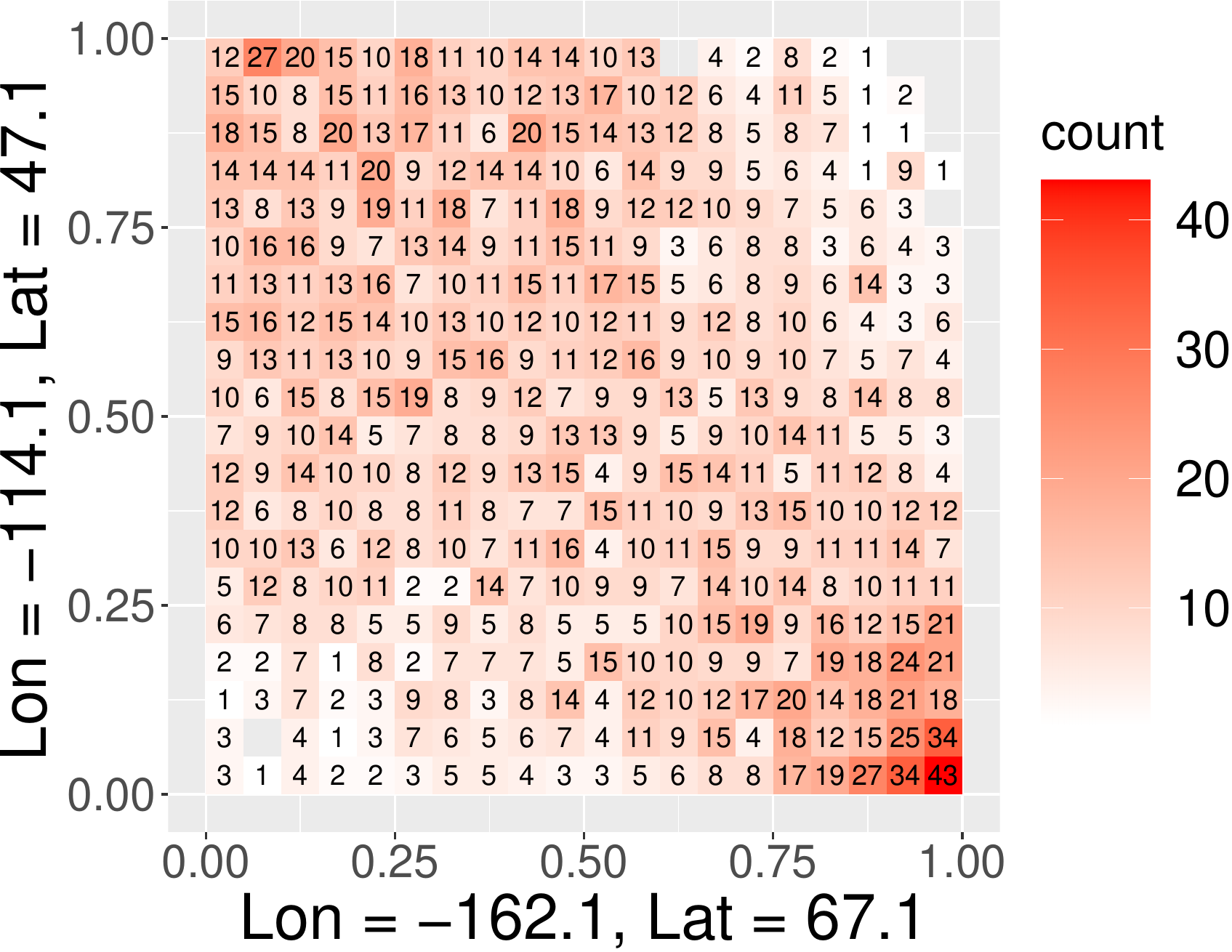}
  \includegraphics[scale=.4]{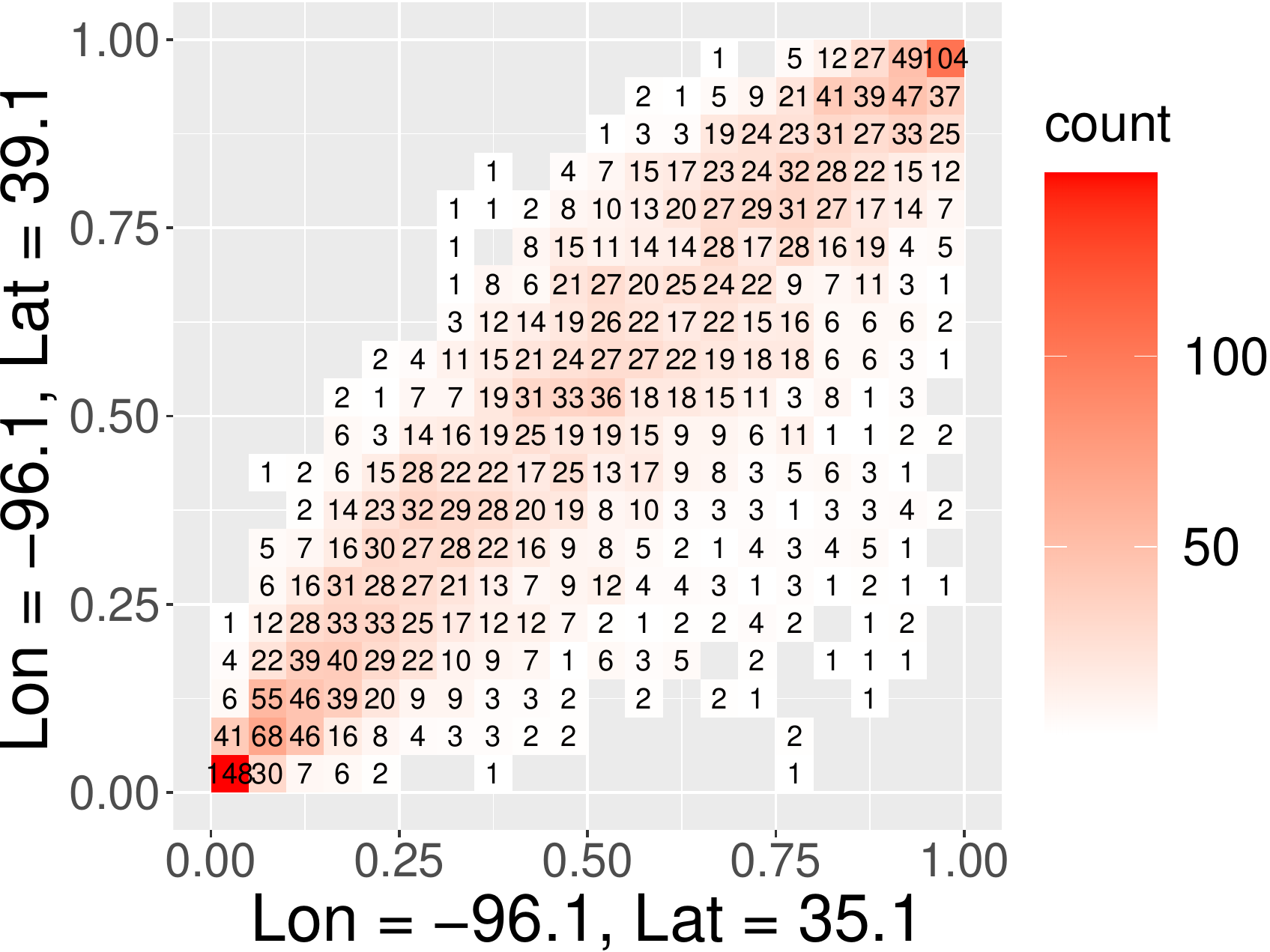}
  \caption{Counts of the number of data points in square bins (on the copula scale). Bin width is 0.05.}
  \label{fig:countmap}
\end{figure}

We show some additional plots of densities corresponding to Main Figure 2 in the paper. 
Supplementary Figure \ref{fig:margscalelogdensity} shows the log copula densities from Main Figure 2 but with axes transformed to the temperature scale (degrees Celsius) instead of the unit square. 
More specifically, the copula density $c$ equals zero outside of the unit square, but here we are plotting the log of $c(F_X(x), F_Y(y))$ as a function of $x$ and $y$ on the temperature scale.
This rescaling to the temperature domain makes various asymmetries in the dependence structure more obvious. 
A notable result is how the blocking extreme in the lower right corner of the left plot in Supplementary Figure \ref{fig:margscalelogdensity} is deformed by this transformation. 
The density remains large when the midlatitude cold extreme temperatures are increased slightly from their lowest values. However, in the opposite case when the Alaska hot extreme temperatures are slightly decreased, the density decreases much faster.
Supplementary Figure \ref{fig:jointlogdensity} shows the logarithm of the joint density $f_{X,Y}(x,y) = c(F_X(x),F_Y(y)) f_X(x) f_Y(y)$ of the two pairs of locations. Here it is harder to discern the tail behaviors, as the marginal densities take relatively small values in the tails of the marginal distribution.
Main Figure 2 may give the impression of strong symmetries within these pairs of locations, but from these two supplementary figures, it is clear that there are many interesting asymmetries in their dependence structures.
Finally, Supplementary Figure \ref{fig:countmap} shows counts of the copula data to which our model is fit.

